\documentclass{pasj01}
\usepackage{url}
\Received{2018/12/11}
\Accepted{2019/3/4}
\Published{$\langle$publication date$\rangle$}

\newcommand{\eprint}[2][]{\url{#2}}

\begin{document}

\title{Ionization age of iron ejecta in the Galactic Type~Ia supernova remnant G306.3$-$0.9}
\author{Makoto Sawada$^{1,2,3}$, Katsuhiro Tachibana$^4$, Hiroyuki Uchida$^4$, Yuta Ito$^3$, Hideaki Matsumura$^{5, 4}$, Aya Bamba$^{6,7,3}$, Takeshi Go Tsuru$^4$, and Takaaki Tanaka$^4$}%
\altaffiltext{1}{X-ray Astrophysics Laboratory, NASA Goddard Space Flight Center, Greenbelt, MD 20771, USA}
\altaffiltext{2}{Department of Physics, University of Maryland Baltimore County, 1000 Hilltop Circle, Baltimore, MD 21250, USA}
\altaffiltext{3}{Department of Physics and Mathematics, Aoyama Gakuin University, 5-10-1 Fuchinobe, Sagamihara, Kanagawa 252-5258, Japan}
\altaffiltext{4}{Department of Physics, Graduate School of Science, Kyoto University, Kitashirakawa Oiwake-cho, Sakyo-ku, Kyoto 606-8502, Japan}
\altaffiltext{5}{Kavli Institute for the Physics and Mathematics of the Universe (WPI), The University of Tokyo Institutes for Advanced Study, The University of Tokyo, 5-1-5 Kashiwanoha, Kashiwa, Chiba 277-8583, Japan}
\altaffiltext{6}{Department of Physics, The University of Tokyo, 7-3-1 Hongo, Bunkyo-ku, Tokyo 113-0033, Japan}
\altaffiltext{7}{Research Center for the Early Universe, School of Science, The University of Tokyo, 7-3-1 Hongo, Bunkyo-ku, Tokyo 113-0033, Japan}
\email{makoto.sawada@nasa.gov}


\maketitle

\begin{abstract}
We present a 190~ks observation of the Galactic supernova remnant (SNR) G306.3$-$0.9 with Suzaku. To study ejecta properties of this possible Type~Ia SNR, the absolute energy scale at the Fe-K band was calibrated to a level of uncertainty less than 10~eV by a cross-calibration with the Hitomi microcalorimeter using the Perseus cluster spectra. This enabled us for the first time to accurately determine the ionization state of the Fe~K$\alpha$ line of this SNR. The ionization timescale ($\tau$) of the Fe ejecta was measured to be $\log_{10} \tau$ (cm$^{-3}$~s) $=$10.24$\pm$0.03, significantly smaller than previous measurements. Marginally detected K$\alpha$ lines of Cr and Mn have consistent ionization timescales with Fe. The global spectrum was well fitted with shocked interstellar matter (ISM) and at least two ejecta components with different ionization timescales for Fe and intermediate mass elements (IME) such as S and Ar. One plausible interpretation of the one-order-of-magnitude shorter timescale of Fe than that of IME ($\log_{10} \tau = $11.17$\pm$0.07) is a chemically stratified structure of ejecta. By comparing the X-ray absorption column to the H\emissiontype{I} distribution decomposed along the line of sight, we refined the distance to be $\sim$20~kpc. The large ISM-to-ejecta shocked mass ratio of $\sim$100 and dynamical timescale of $\sim$6~kyr place the SNR in the late Sedov phase. These properties are consistent with a stratified ejecta structure that has survived the mixing processes expected in an evolved supernova remnant.
\end{abstract}

\section{Introduction}

Type~Ia supernovae (SNe) play a crucial role in the chemical evolution of galaxies because they are the primary source of Fe-peak elements. 
One of the clues for investigating the physics of Type~Ia SNe is X-ray spectra of supernova remnants (SNRs).
Because of a reasonably symmetric explosion in a uniform surrounding medium, young Type~Ia SNRs generally have symmetrical morphologies \citep{2009ApJ...706L.106L}, which enable us to observationally investigate the stratified structure of heavy elements directly reflecting the explosion mechanism. 
\citet{2010ApJ...725..894H} found significantly slower expansion of Fe than those of Si, S, and Ar in Tycho's SNR based on the Doppler measurements and interpreted it to be the evidence for segregation of Fe ejecta to the inner region, which is naturally expected from the Type~Ia nucleosynthesis.
Another example is SN\,1006, where \citet{2008PASJ...60S.141Y} detected low-ionization Fe K-shell emission from the center of the remnant. 
Element-dependent ionization timescales suggested that the observed Fe has been heated by the reverse shock more recently than the lighter elements. 
\citet{2010A&A...519A..11K} revealed the layering of the elements in SNR 0519$-$69.0 with radial emissivity profiles using Chandra images.
There are also some Type~Ia SNRs that appear not so nicely layered. 3C\,397 \citep{2015ApJ...801L..31Y} and W\,49B \citep{2018A&A...615A.150Z} are such examples, although their type identifications are still controversial.
As demonstrated by these previous studies, emission lines from the shock-heated ejecta can be used as tracers of the distribution of the nuclear-burning products. The Fe K-shell emission is of particular importance among the available thermal X-ray lines, as it traces a product at the end point of the Ia nucleosynthesis, and thus has the most contrasted distribution compared to those of intermediate-mass elements (IME: Si, S, Ar, and Ca). 

The stratified structure may be modified by hydrodynamical instabilities as an SNR evolves (\cite{1992ApJ...392..118C}, \cite{2001ApJ...549.1119W}).
This is particularly the case if the circumstellar environment largely deviates from the uniform and homogeneous distribution. One example is a bubble made by accretion wind from a companion star. \citet{2007ApJ...662..472B} showed that the interaction between ejecta and a dense radiative bubble formed by low-speed accretion wind (model HP3u1e6) enhances ionization of both Si and Fe compared to the uniform ISM case, and eventually equalizes the average ionization timescales of these elements at a few kyr. 
\citet{2012ApJ...745...75G} demonstrated another possibility with three-dimensional simulations that a hole in ejecta carved by a companion star affects the long-term evolution of an SNR. In this model, large-scale hydrodynamical instabilities at the edge of the hole trigger the intrusion of material into the hole during the Sedov phase. Such an effect may cause an efficient mixing of ejecta. Aside from the SNR evolutions, theoretical studies of the single-degenerate scenario even suggest the possibility that ejecta may be completely mixed at the time of explosion by multiple igniting bubbles (\cite{2005A&A...430..585G}, \cite{2005ApJ...624..198B}). 
However, there has been only a limited number of samples to observationally investigate these possibilities. Only 14 samples have been identified as Type~Ia SNRs or candidates with clear detection of the Fe~K$\alpha$ emission (\cite{2014ApJ...785L..27Y}, \cite{2017A&A...597A..65M}), and eight among them have been examined for elemental dependence of ionization timescales with the latest generation of X-ray CCD spectrometers. Most of them, including  G344.7$-$0.1 \citep{2012ApJ...749..137Y}, SN\,1006 \citep{2013ApJ...771...56U}, Kepler's SNR, Tycho's SNR, and SNR\,0509$-$6.5 \citep{2015ApJ...808...49K}, have relatively low ionization timescales ($\log_{10}\tau<$11 for IME and $\log_{10}\tau<$10 for Fe), suggesting the samples are biased toward young Ia SNRs. Therefore it is still unknown when and how the stratified structure of a Type~Ia SNR is distorted by mixing, in particular owing to external influences.

The SNR G306.3$-$0.9 was discovered with the Swift Galactic plane survey \citep{2011ATel.3415....1M}. 
\citet{2013ApJ...766..112R} conducted a 5~ks follow-up observation with Chandra and revealed the distorted morphology with brightening toward the southwest. The X-ray spectrum showed emission lines from highly ionized atoms of Mg, Si, S, and Ar. The spectrum was described by a non-equilibrium ionization plasma with a temperature below 1~keV. The small extension of $\sim$4~pc radius assuming a distance of 8~kpc and the Sedov model makes the remnant a young Galactic SNR with an age of about 2.5~kyr. 

\citet{2016A&A...592A.125C} (hereafter CO16) reported the results of longer-exposure observations with XMM-Newton and Chandra. With $\sim$50~ks observation for each, they performed spatially resolved X-ray spectroscopy and discovered K$\alpha$ emission lines of Ca and Fe from the central and southwestern parts. To explain the existence of these lines, they revised a view of the X-ray spectra from a single-temperature plasma to two-temperature plasma consisting of ISM and single ejecta components with temperatures of $\sim$0.2 and 1--2~keV, respectively. The Fe~K$\alpha$ line had a centroid of 6.52$\pm$0.01~keV. They concluded that G306.3$-$0.9 is a remnant of a Type~Ia SN based on both the relatively low-density environment suggested by the Fe~K$\alpha$ line \citep{2014ApJ...785L..27Y} and the relatively low ellipticity of the X-ray morphology derived from the multipole-moment image analysis \citep{2011ApJ...732..114L}. The ionization timescales were measured at $\log_{10}\tau=$10.5--11.0. Thus, the SNR is a good candidate to study the ejecta distribution in a relatively later phase in thermal evolution.
Recently, \citet{2017MNRAS.466.3434S} (hereafter SE17) reported results of an X-ray observation with Suzaku \citep{2007PASJ...59S...1M} of G306.3$-$0.9. The X-ray spectra were reproduced with a similar model to CO16 consisting of the ISM and ejecta components, while the temperatures were significantly different ($\approx$0.6~keV for ISM and $\approx$3~keV for ejecta). The Fe~K$\alpha$ centroid was measured at a slightly lower value, 6.50$\pm$0.01~keV. 
In these studies, the ejecta emission from Si to Fe was reproduced by a single-component plasma, implying the common ionization timescales for the range of elements. 
A natural interpretation to this is that the stratification has been significantly reduced by mixing, which is clearly different from what we observe in other Ia remnants. Because the characterization of the stratified ejecta largely relies on the measurement of ionization timescale of Fe, the accuracy of the Fe~K$\alpha$ centroid should be a key factor. The recent two studies reported Fe~K$\alpha$ centroids that disagree at the 1.0$\sigma$ level but agree at the 2.0$\sigma$ level. This raises the possibility that a revised data analysis with the latest, most accurate calibration might resolve this apparent discrepancy.

In this paper, we present a re-analysis result of the X-ray spectrum of the entire remnant with the X-ray Imaging Spectrometer (XIS: \cite{2007PASJ...59S..23K}) onboard Suzaku. The deepest exposure of about 190~ks together with cross-calibrated energy scale around 7~keV with the X-ray micro-calorimeter (SXS: \cite{2014SPIE.9144E..2AM}) onboard Hitomi \citep{2016SPIE.9905E..0UT} allows us to study the nature of the Fe ejecta accurately. We discover that the Fe ejecta has significantly smaller ionization timescale than IME, which recovers a consistent view of the chemically stratified ejecta structure of a Type~Ia SNR. 

\section{Observation and Data Reduction}\label{sec:reduction}

\begin{figure*}[!htbp]
\begin{center}
\includegraphics[width=17cm]{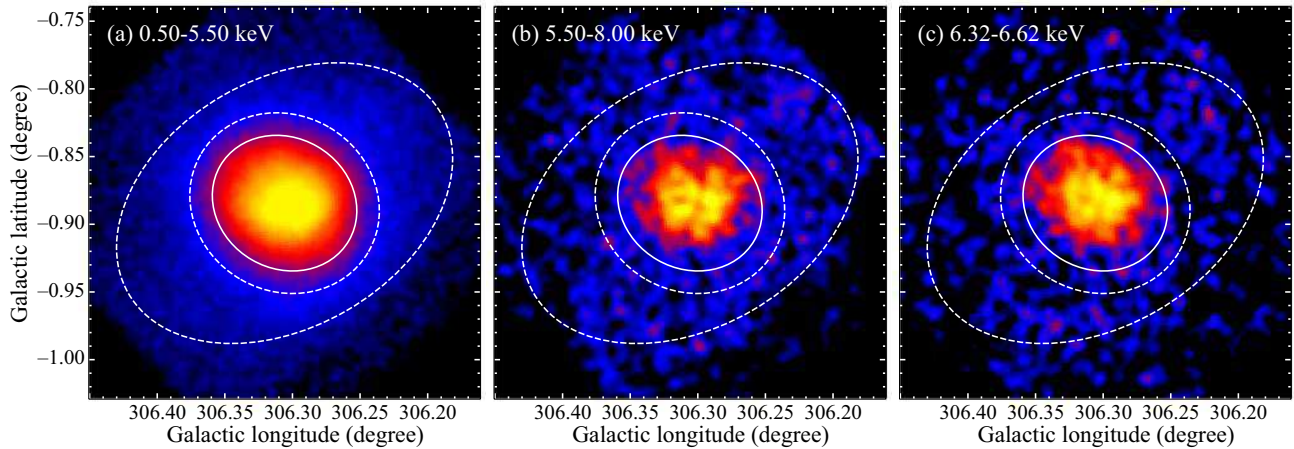}
\end{center}
\caption{Band-limited images with XIS\,0+1+3 in the logarithmic scale: (a) 0.50--5.50~keV, (b) 5.50--8.00~keV, and (c) 6.32--6.62~keV tracing Fe~K$\alpha$. The overlaid ellipses are the source (solid) and background (dashed) extraction regions.}\label{fig:images}
\end{figure*}

G306.3$-$0.9 was observed with the XIS on 2014 August 20--25 (sequence number: 509072010) 
with an aim point of (R.A., Decl.)$_{\rm J2000.0}$ = (\timeform{13h21m54.072s}, \timeform{-63D33'36.72''}). 
The entire remnant was covered by a \timeform{17'.8}$\times$\timeform{17'.8} field of view (FOV) of the XIS.

The XIS consists of four X-ray CCD cameras, placed at the focal planes of the 
X-Ray Telescope (XRT: \cite{2007PASJ...59S...9S}). Three (XIS\,0, 2, 3) are front-illuminated (FI) sensors, 
while the other (XIS\,1) is a back-illuminated (BI) one. The entire area of XIS\,2 and a quarter edge of XIS\,0 
have not been functional since 2006 November and 2009 June, respectively. 
The angular resolutions of XRTs are 1\arcmin.9--2\arcmin.3 in the half-power diameter (HPD)
with almost no dependence on X-ray energies and off-axis angles within $\sim 10$\%. 
The total effective area of the operational three sensors is 480~cm$^2$ at 8 keV. 
The low-Earth orbit at an altitude of $\approx$550~km enables the low and stable background environment. 

The XIS was operated with the full-window and normal-clocking mode. 
The spaced-row charge injection was applied to mitigate the degradations in the energy gain and resolution. 
The data reduction was done with the software package HEASoft version 6.22 and the pipeline processing 
version 3.0. 
With this processing, the latest and final Suzaku calibration database released on 2016 April 01 
was applied to register energies of X-ray events using the method described by 
\citet{2009PASJ...61S...9U} and \citet{2012AIPC.1427..245S}.
We removed events during the South Atlantic Anomaly passages, the Earth elevation angles below 5\arcdeg, 
the Earth daytime elevation angles below 20\arcdeg. The exposure after the reduction is 190~ks. 
We eliminated events of flickering pixels with noisy pixel maps\footnote{\label{fnm:1}See $\langle$https://heasarc.gsfc.nasa.gov/docs/suzaku/analysis/xisnxbnew.html$\rangle$.}. We further removed their adjacent neighbors as these also contributed to degrading spectral performance for the present data. These maskings reduced the event numbers to 85\%, 45\%, and 89\% for XIS\,0, 1, and 3, respectively. The higher rejection rate for XIS\,1 is partly attributable to the intrinsically higher fraction of noisy pixels than other sensors, which is essentially because of the different manufacturing processes (i.e., BI vs. FI).

Response files were generated by xisrmfgen and xissimarfgen \citep{2007PASJ...59S.113I} and 
Non--X-ray Background (NXB) events were 
retrieved by xisnxbgen \citep{2008PASJ...60S..11T} with the same version of the calibration database. 
In the calculation of the effective area, an extended source with uniform X-ray brightness was assumed. 
Losses of effective area due to the removal of the noisy pixels and the neighbors were taken into account 
by the method provided by the XIS team\footnotemark[\ref{fnm:1}]{}.

\section{Analysis and Results}

\begin{figure}[!htbp]
\begin{center}
\includegraphics[width=8cm]{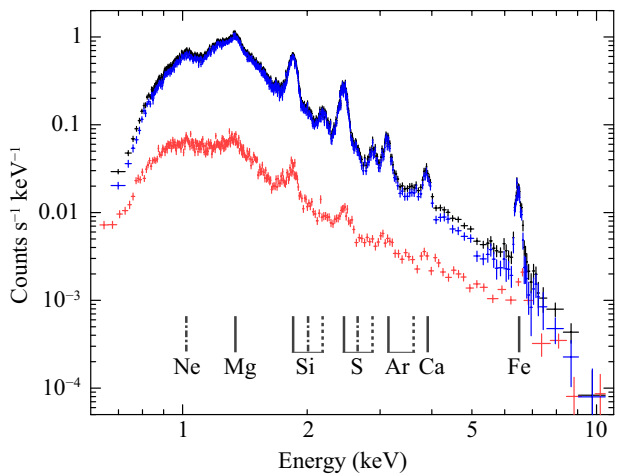}
\end{center}
\caption{Spectra with the FI sensors (XIS\,0 + 3) from the source (black) and background (red) regions overlaid with the result of the background subtraction (blue). The vertical bars show the energies of the detected K-shell emission lines: the solid bars for $n=$1--2 lines from He-like ions (He$\alpha$) or lower ionization ions, the dashed bars for $n=$1--3 lines from He-like ions (He$\beta$), and the dot-dashed bars for $n=$1--2 lines from H-like ions (Ly$\alpha$).}
\label{fig:bgsub}
\end{figure}

\subsection{Images}

Figure~\ref{fig:images} shows band-limited XIS images of G306.3$-$0.9. Thanks to the low NXB level of the XIS, we obtained the hard-band image above 5~keV for the first time. The spatial distribution was found to be nearly identical between the hard band (b) and the Fe~K$\alpha$ band (c), indicating the thermal origin of the continuum emission above 5~keV.

The angular size of the remnant, \timeform{110"} in radius, is comparable to the HPD of the XRT (\timeform{1.9'}--\timeform{2.3'}: \cite{2007PASJ...59S...9S}). This makes it difficult to analyze spatially resolved spectra. Actually, SE17 found no significant difference in spectral shapes (absorption columns and electron temperatures) when spectra from inner and rim regions were compared. Thus, we decided to analyze the emission from the entire remnant, and determined the source and background regions for the spectral analysis as indicated with solid and dashed ellipses, respectively, in figure~\ref{fig:images}. 

\subsection{Spectral analysis}\label{sec:spectra}

For spectral analysis, we used Xspec version 12.9.1p with AtomDB version 3.0.9 unless otherwise noted. Charge-state distributions were calculated based on ionization and recombination coefficients compiled by \citet{2009ApJ...691.1540B} for emission models. Photo-electric cross sections by \citet{1992ApJ...400..699B} were used for absorption models. The solar abundances of \citet{1989GeCoA..53..197A} were used for the both types of models. Uncertainties are quoted at 90\% confidence intervals while error bars in figures show 1-$\sigma$ intervals unless otherwise noted. 

Figure~\ref{fig:bgsub} shows the FI spectra extracted from the source and background regions after energy-scale correction as described below. The excess emission above the background is significant up to $\approx$9~keV. The detections of the K-shell lines of Ca and Fe recently reported by CO16 were confirmed with the XIS spectra with rich statistics. 

In the model fitting analysis, energy bins near the Si-K edge were ignored because of the limited calibration accuracy of the edge structure\footnote{See $\langle$https://heasarc.gsfc.nasa.gov/docs/suzaku/aehp\_data\_analysis.html$\rangle$.}. Different energy ranges were defined and eliminated for FIs (1.70--1.90~keV) and BI (1.75--1.85~keV) to mitigate a normalization inconsistency at Si He$\alpha$ originating from the same issue. For the BI data, bins above 7~keV were also ignored because of large systematic uncertainty in an X-ray flux due to the higher NXB rate. 

\subsubsection{Cross-calibration of energy scale using Hitomi SXS}\label{sec:energycal}

Accurate energy-scale in 6--7 keV is required to determine the ionization timescale of Fe ejecta using the Fe~K$\alpha$ energy centroid based on CCD spectra. The inner two segments (B and C: \cite{2007PASJ...59S..23K}; the Suzaku technical description\footnote{See $\langle$https://heasarc.gsfc.nasa.gov/docs/suzaku/prop\_tools/suzaku\_td/suzaku\_td.html$\rangle$.}) of the XIS at this energy band have been calibrated with Fe He$\alpha$ from the Perseus cluster. A 20~ks observation of the calibration target was conducted one week after G306.3$-$0.9 (sequence number: 109005010). An unprecedented high-resolution observation of this cluster was recently carried out with the SXS onboard Hitomi (e.g., \cite{2016Natur.535..117H}, \cite{2017Natur.551..478H}). The energy resolution of $\approx$5~eV at FWHM and the absolute energy accuracy of $\approx$1~eV at the Fe-K band \citep{2016SPIE.9905E..3UL} make this observation the best reference ever. Therefore, we examined the XIS energy scale in this band against the SXS spectrum. 

Using the closest XIS observation of the Perseus cluster, being conducted and processed in the same manner as that of G306.3$-$0.9, we extracted the FI spectrum from the SXS FOV. The SXS spectrum of the Fe He$\alpha$ is characterized by a $\approx$4~keV CIE plasma with a redshift of 0.01756 (e.g., \cite{2016Natur.535..117H}), but is better reproduced by a three-temperature CIE model (an ``improved model'' by \cite{2018PASJ...70...12H}), thus we applied the latter model as the reference spectrum. As this model was based on SPEX version 3.03, we imported it to Xspec by using \texttt{flx2tab} and compared to the XIS spectrum in 6.38--6.68~keV. An energy shift was then added as a free parameter, which was measured to be $+$4.2$\pm$6.6~eV. This result shows that the FI energy scale at the Fe-K band is highly reliable at a level of $\lesssim$10~eV near the observation period of G306.3$-$0.9. The BI energy scale in this band, on the other hand, showed a few 10~eV deviation from both the FI and SXS energies, and was corrected to match the SXS. 
We further checked possible energy shifts in the lower energy bands below 4~keV using theoretical values of the He$\alpha$ and Ly$\alpha$ lines of Si, S, Ar, and Ca, but found no significant shifts (i.e., shifts larger than $\approx$10~eV), either in the FI or BI spectrum. 

\subsubsection{Single ejecta models}\label{sec:singleejecta}

\begin{figure*}[htbp]
\begin{center}
\includegraphics[width=8cm]{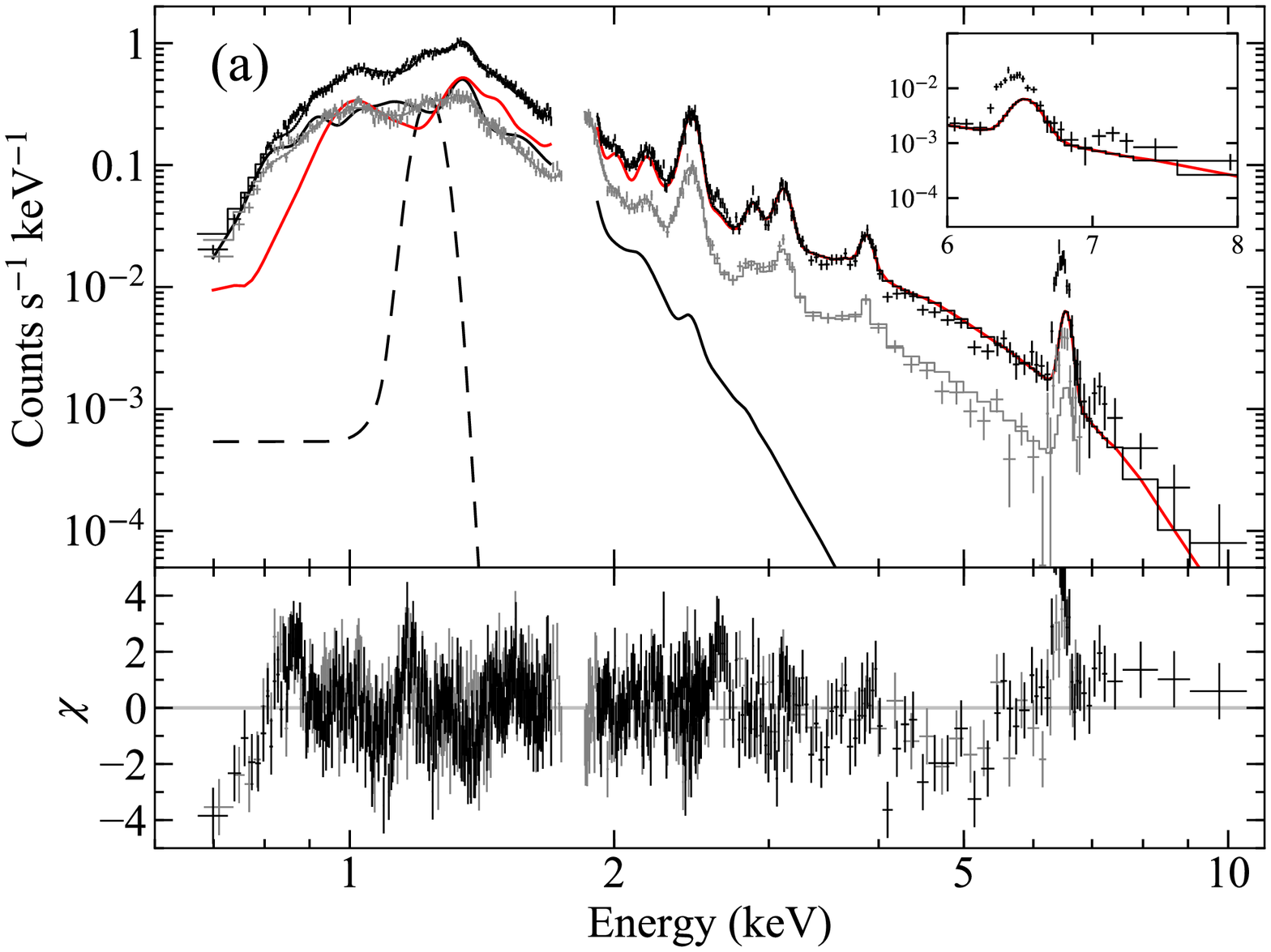}
\includegraphics[width=8cm]{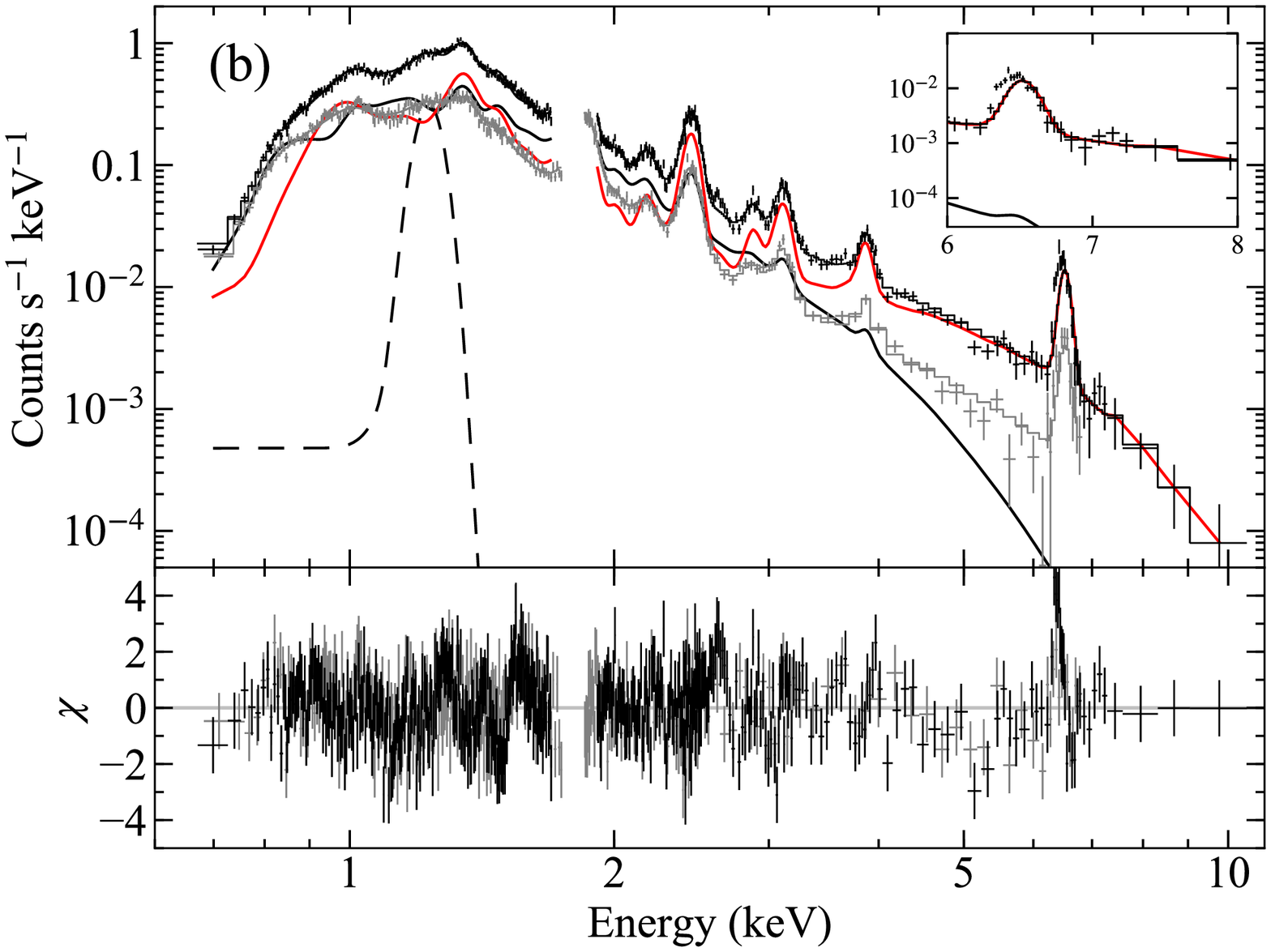}\\
\includegraphics[width=8cm]{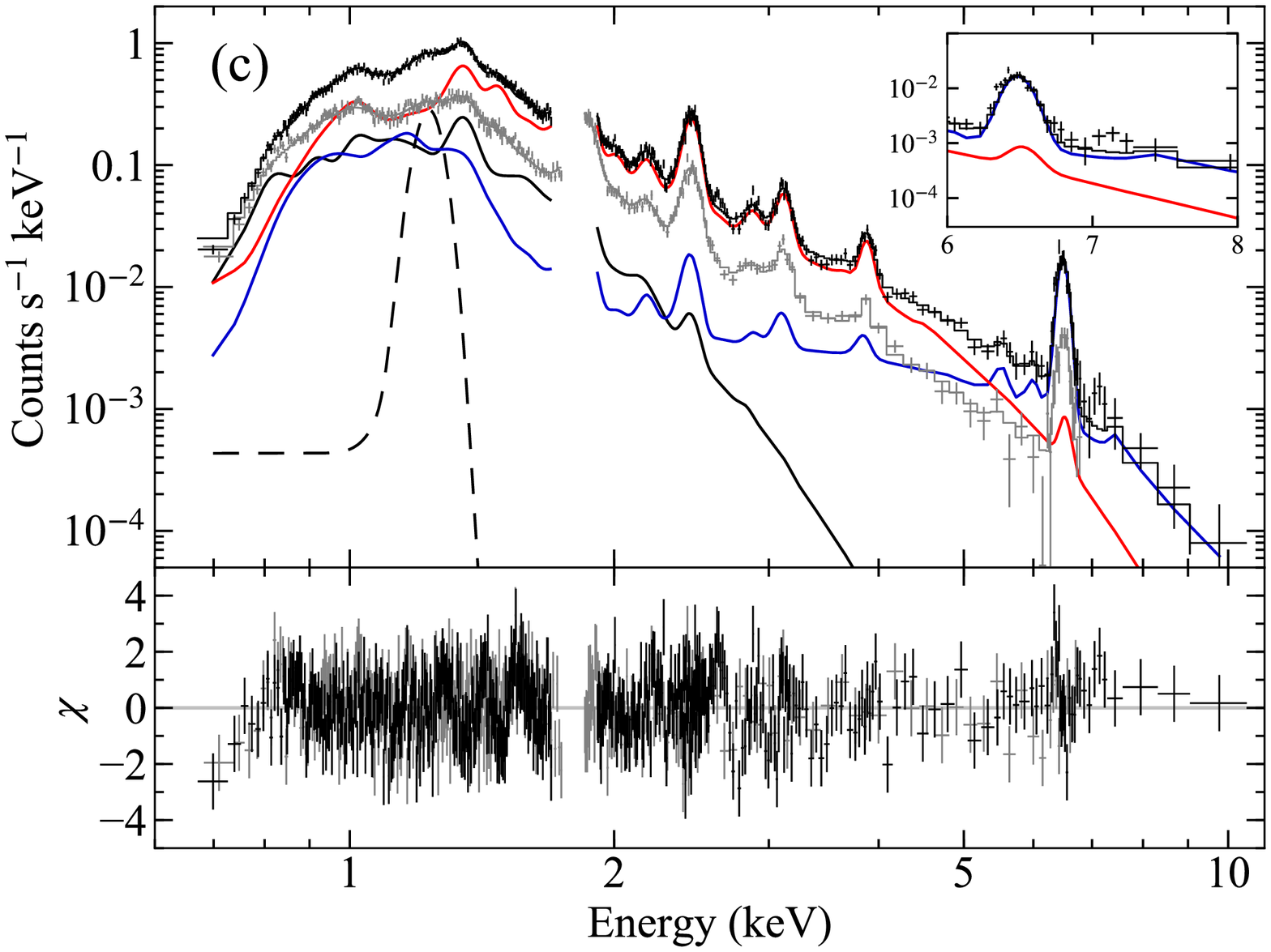}\\
\includegraphics[width=8cm]{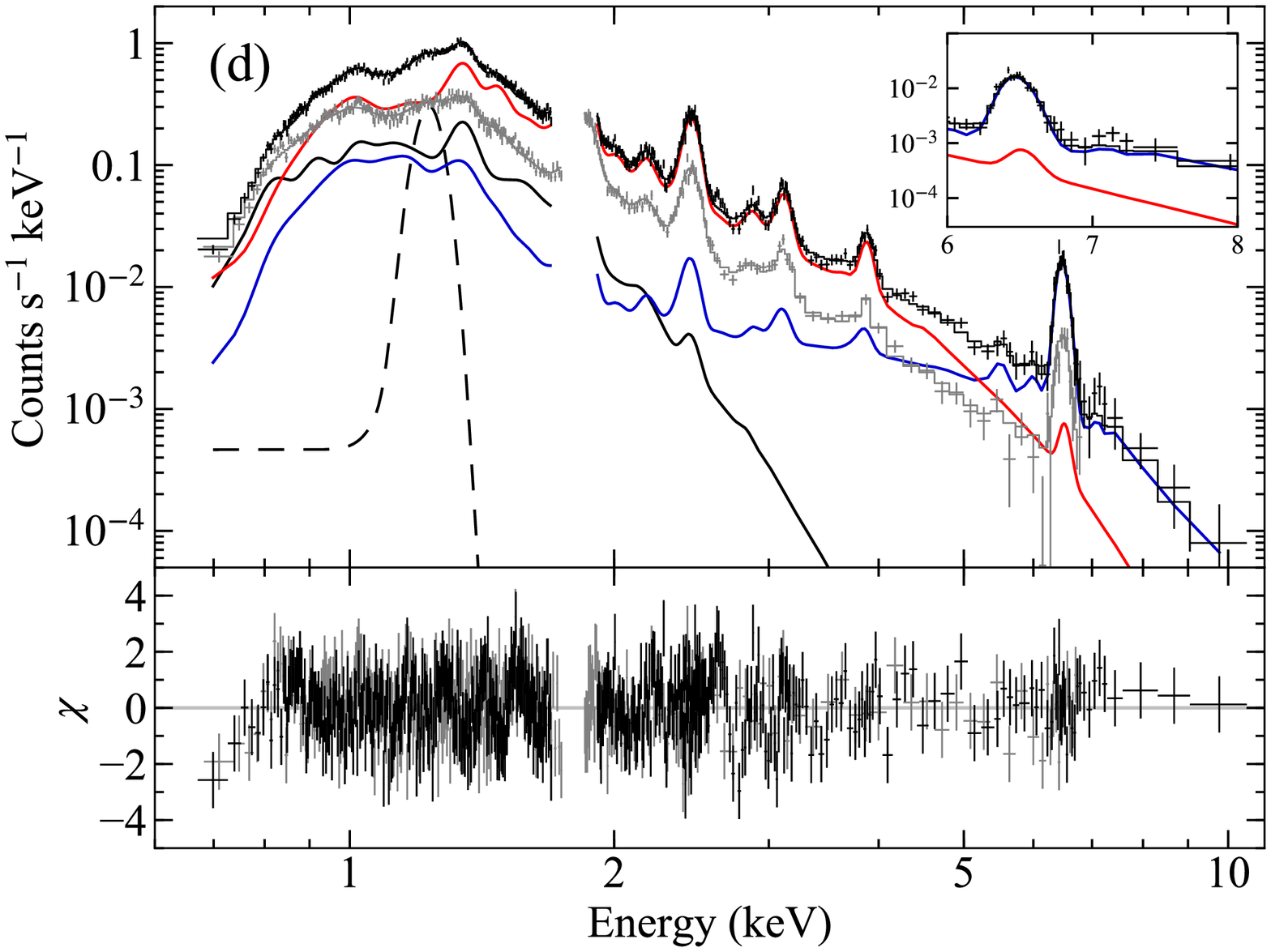}
\includegraphics[width=8cm]{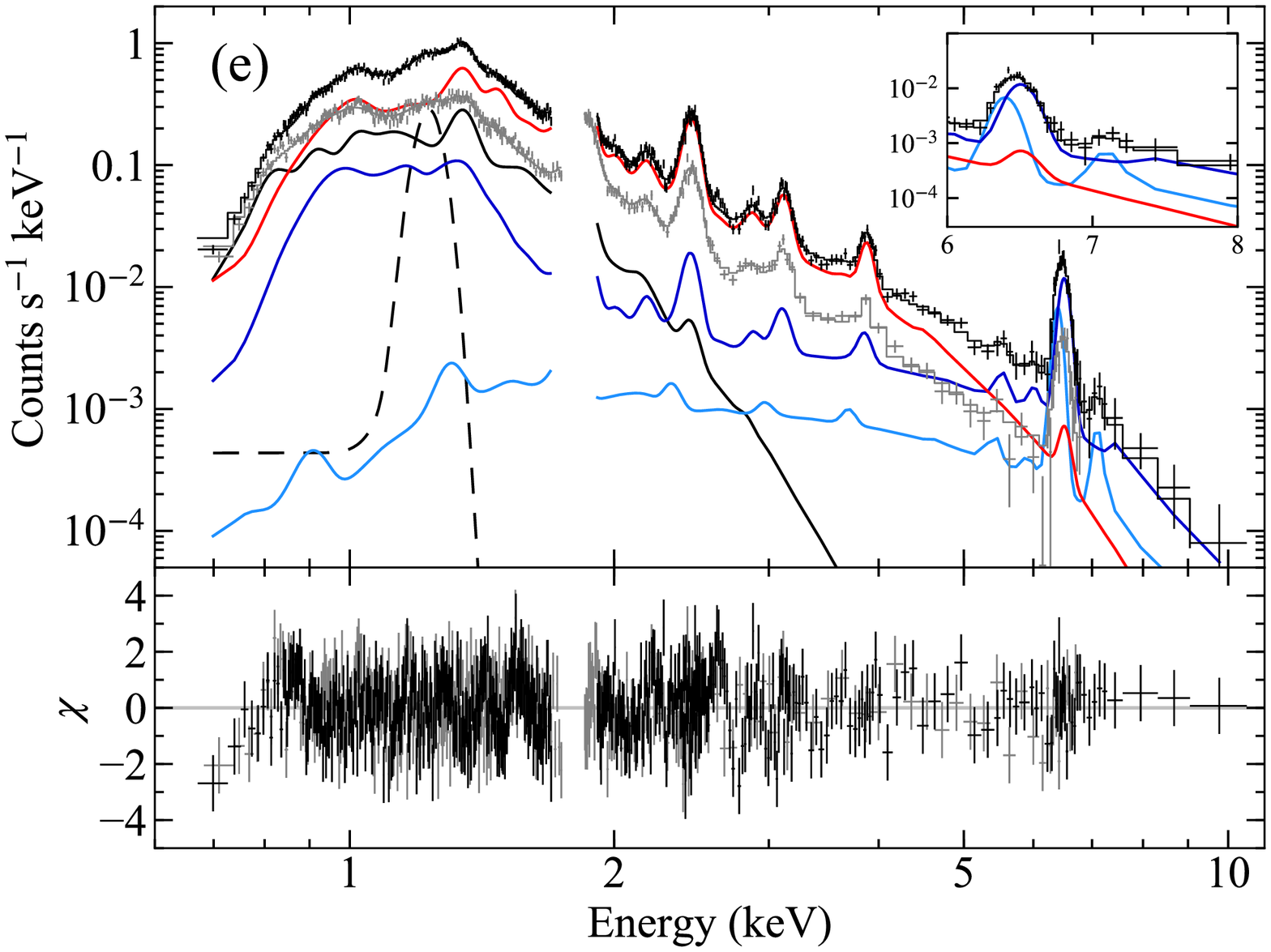}
\end{center}
\caption{Background-subtracted spectra (black crosses for FI and gray ones for BI) with the best-fit models (histograms). Individual panels show results of different ejecta modelling: single ejecta models (a and b: subsection~\ref{sec:singleejecta}) and multi-$\tau$ ejecta models (c--e: subsection~\ref{sec:multitau}). The single ejecta models are those with the CO16-like parameters (a) and SE17-like ones (b). The multi-$\tau$ ejecta models assume different Fe-dominant ejecta conditions: single-$\tau$ Fe ejecta (c), continuous-$\tau$ Fe ejecta under a plane-parallel shock condition (d), and discrete-$\tau$ Fe ejecta characterized by two-$\tau$ components (e). The black dashed and dotted curves are the common components and show \texttt{apec} for ISM and Gaussian for the Fe-L line, respectively. The solid curves with red, blue, and light blue are the ejecta components and show Ejecta\,1, 2, and 3, respectively (table~\ref{tab:parameters}). The insets show closed-up view of the FI spectra in the Fe-K band. The labels on individual subfigures correspond to the model names in the text and table~\ref{tab:parameters}.}
\label{fig:specfit}
\end{figure*}

We performed spectral fittings with models of optically thin thermal plasma attenuated by interstellar absorption. Chemical abundances ($Z$) of heavy elements with line detections (Ne, Mg, Si, S, Ar, Ca, and Fe), as well as electron temperatures ($kT_{\rm e}$), ionization timescales ($\tau$), emission measures ($EM$), and hydrogen-equivalent absorption column density ($N_{\rm H}$) were allowed to be free in the fittings. To start with, we tried single-temperature plasma models either at collisional ionization equilibrium (CIE) or non-equilibrium ionization (NEI), but obtained poor fits ($\chi^2/\rm{dof} \approx 4$) with large positive residuals above $\approx$ 4~keV, which is consistent with the result studies by CO16 and SE17. 

Next, we tried two-temperature models. Even for the best fits, line-like residuals at $\approx$1.23~keV were always found. This is likely due to known flux deficits of Fe-L lines from highly excited ions (n$\ge$5: \cite{2000ApJ...530..387B}), and thus we thereafter added a Gaussian at a fixed energy of 1.23~keV. This is a crude approximation because actual Fe-L lines are ensembles of many different transitions, but it turned out to be sufficient for the present data given the energy resolution and statistics. Similar treatments are found in the analysis of other SNRs (e.g., \cite{2012PASJ...64...81S}, \cite{2016PASJ...68S...8S}, and \cite{2016MNRAS.462.3845K}). We found that combinations of a low-temperature CIE with the solar abundance and a high-temperature NEI with variable abundance gave relatively good fits. The CIE and NEI components can be attributed to shocked ISM and ejecta, respectively. We made several trial fits with different initial values of the fitting parameters and found that there were two local minima. One has relatively low temperatures of 0.2~keV for ISM and 1.6 keV for ejecta, while the other has higher values of 0.6~keV and 2.8~keV. The two sets of the fitted parameter values are similar to those reported by CO16 and SE17, respectively. 
Both the models explain ejecta emission as a single NEI component, and thus we refer to these as the single-ejecta models (a) and (b). 
The best-fit parameters are listed in the columns (a) and (b) of table~\ref{tab:parameters}. 
Although the previous studies reported that a single-ejecta model was enough to reproduce the full-band spectrum of G306.3$-$0.9, notable ``N''-shaped residuals are seen around the Fe~K$\alpha$ line (figure~\ref{fig:specfit}a and b). 
It should also be pointed out that, even with the different calibration databases (section~\ref{sec:band-limited}), the ``N''-shaped residuals were seen in the previous Suzaku fitting results (see figure~2 in SE17). These facts obviously show that the applied single-ejecta models predict higher centroids than the XIS data. 

\subsubsection{Band-limited fittings}\label{sec:band-limited}

\begin{figure}[!htbp]
\begin{center}
\includegraphics[width=8cm]{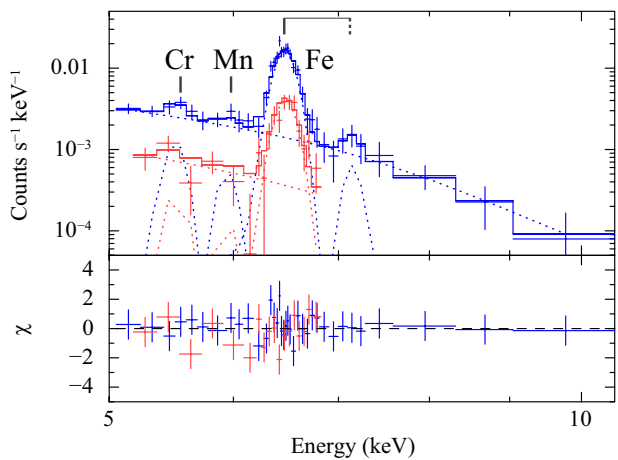}
\end{center}
\caption{Fe-K spectra (blue crosses for FI and red ones for BI) fitted with Gaussians plus bremsstrahlung (histograms). Individual contributions from emission components are shown with the dotted curves.}
\label{fig:Fefit}
\end{figure}

The failures in reproducing the Fe~K$\alpha$ centroid with the single ejecta models indicate the difference of ionization timescales between IME and Fe. Because the ionization timescale of Fe ejecta is obtained from the Fe~K$\alpha$ centroid, we performed a band-limited fitting above 5~keV and derived the centroid value. An empirical model consisting of Gaussians and bremsstrahlung was applied. Figure~\ref{fig:Fefit} shows a close-up view of the 5.0--10.5~keV band with the best-fit model. The centroid and width of the Gaussian for Fe~K$\alpha$ were derived to be 6477.2$^{+7.5}_{-9.5}$~eV and 57.8$^{+12.1}_{-14.1}$~eV, respectively. There are also hints of other emission lines particularly in the FI data. The line centroids were measured at 5.51$\pm$0.08, $\approx$5.94, 7.15$^{+0.24}_{-0.14}$~keV, which indicate that these are likely the Cr~K$\alpha$, Mn~K$\alpha$, and Fe~K$\beta$ lines, respectively. The significances are 2.3, 1.1, and 1.7$\sigma$, respectively, requiring a future confirmation. The relative intensities to the Fe~K$\alpha$ line are 5.7$\pm$4.0\%, 2.6 ($<$6.6)\%, and 4.4$\pm$4.2\%, respectively. The bremsstrahlung temperature was measured to be 4.1$^{+4.3}_{-1.6}$~keV. 

\begin{figure}[!htbp]
\begin{center}
\includegraphics[width=8cm]{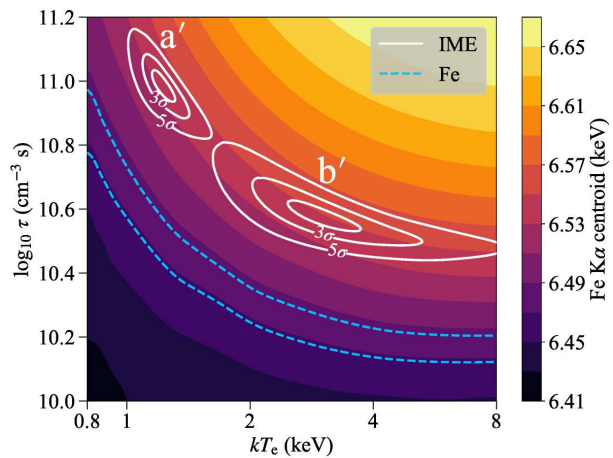}
\end{center}
\caption{Constraints on the NEI parameters ($kT_{\rm e}$ and $\tau$) with the band-limited fittings. The two sets of white contours (a$^{\prime}$ and b$^{\prime}$) show the results of the IME-dominant band in 1.85--4.3~keV based on the two single-ejecta models (a and b) as initial and fixed parameters. The levels are at the 90\%, 3$\sigma$, and 5$\sigma$ intervals. The light blue dashed curves show the constraint by the Fe~K$\alpha$ centroid at the 90\% interval. The colored filled contours show the expected Fe~K$\alpha$ centroids for various combinations of the NEI parameters.}
\label{fig:contours}
\end{figure}

The measured centroid of Fe~K$\alpha$ is significantly lower than the previous results (6.52$\pm$0.01~keV by CO16 and 6.50$\pm$0.01~keV by SE17), and this is the reason why the present spectra cannot be reproduced by the single-ejecta models. We note that a part of the discrepancy between the two Suzaku results is likely due to a difference in the energy scale calibration; we used the latest and final calibration database to calculate the energy scale, which was further tested against and matched to the SXS. 
We then derived Fe K$\alpha$ centroids for various sets of the NEI parameters ($kT_{\rm e}$ and $\tau$) from calculated spectra, as shown with the colored filled contours in figure~\ref{fig:contours}. Using this, we mapped the measured Fe K$\alpha$ centroid into the NEI parameter space (dashed light blue contours in figure~\ref{fig:contours}).

Separately, we can constrain possible combinations of the NEI parameters for IME by fitting the soft-band spectra in the 1.85--4.3~keV band. In this analysis, we started from the best-fit parameters with the full-band fittings (section~\ref{sec:singleejecta}). We then fixed parameters which do not have major spectral features in 1.85--4.3 keV. These are $N_{\rm H}$, $kT_{\rm e}$ and $EM$ of the ISM component (\texttt{apec}), $F_{\rm X}$ of the Gaussian, and $Z$ for Ne, Mg, Fe and Ni of the ejecta component (\texttt{vvnei}). Leaving other parameters free, we derived a two-dimensional confidence limit on $kT_{\rm e}$ and $\tau$ of the ejecta using the \texttt{steppar} command in Xspec. As we obtained two local minima in the full-band fitting with single-ejecta models (a and b), we repeated the same analysis for each of them. The results, which we refer to as a$^{\prime}$ and b$^{\prime}$, are shown in figure~\ref{fig:contours} with two sets of white contours. 

The two confidence limits (a$^{\prime}$ and b$^{\prime}$) show no overlap in the parameter space within 5$\sigma$, but the inferred Fe~K$\alpha$ centroids are similar to each other ($\approx$6.53--6.55~keV) and are significantly higher than that actually measured at Fe K$\alpha$ ($\approx$6.48~keV). This discrepancy already implies different ionization timescales between Fe and IME. The only case in which Fe and IME have the consistent ionization timescales is that $kT_{\rm e}$ of Fe is 1~keV or lower. This possibility can safely be ruled out by the lower limit of 2.5~keV obtained in the hard-band fitting. This sets the upper-limit on the ionization timescale of the Fe ejecta to be $\log_{10} \tau \approx$10.3, suggesting lower $\tau$ value of the Fe ejecta than IME. Below, we further examine this with the full-band analysis.

\begin{table*}[!htdp]
\caption{Best-fit parameters for the single-ejecta and multi-$\tau$ ejecta models in the full-band fittings.}
\begin{center}
\begin{tabular}{lllccccc}
\hline
Component & \multicolumn{2}{l}{Parameter} & \multicolumn{2}{c}{\dotfill Single ejecta\dotfill} & \multicolumn{3}{c}{\dotfill Multi-$\tau$ ejecta\dotfill} \\ 
& & & CO16 like & SE17 like & Single-$\tau$ Fe & \multicolumn{2}{c}{\dotfill Multi-$\tau$ Fe\dotfill}\\ 
& & & & & & Continuous & Discrete \\
& & & (a) & (b) & (c) & (d) & (e) \\
\hline
\multicolumn{8}{c}{\dotfill Common components\dotfill} \\
\texttt{phabs} & $N_{\rm H}$ & ($10^{22}$ cm$^{-2}$) &  $1.37^{+0.02}_{-0.03}$ &           $1.21\pm0.01$ &            $1.22\pm0.04$ &            $1.24\pm0.03$ &   $1.27^{+0.04}_{-0.05}$ \\
\texttt{apec} & $kT_{\rm e}$ & (keV) &           $0.23\pm0.01$ &           $0.61\pm0.01$ &   $0.28^{+0.06}_{-0.03}$ &   $0.27^{+0.02}_{-0.03}$ &            $0.27\pm0.03$ \\
     & $EM$ & ($10^{10}$ cm$^{-5}$) &    $1609^{+243}_{-224}$ &               $146\pm7$ &      $289^{+149}_{-139}$ &      $324^{+262}_{-107}$ &      $414^{+269}_{-159}$ \\
Gaussian & $F_{\rm X}$ & ($10^{-4}$ cm$^{-2}$ s$^{-1}$) &  $7.28^{+0.30}_{-0.32}$ &           $5.02\pm0.23$ &   $4.65^{+0.39}_{-0.40}$ &   $5.12^{+0.38}_{-0.37}$ &   $5.01^{+0.56}_{-0.42}$ \\
\multicolumn{8}{c}{\dotfill Ejecta components\dotfill} \\
Ejecta\,1 & $kT_{\rm e}$ & (keV) &  $1.57^{+0.07}_{-0.06}$ &  $2.81^{+0.12}_{-0.13}$ &   $0.99^{+0.07}_{-0.06}$ &            $0.93\pm0.05$ &   $0.92^{+0.09}_{-0.07}$ \\
(\texttt{vvnei})     & $Z$ & ... Ne &                 $<0.82$ &         $0.00^{<+0.66}$ &   $0.62^{+0.29}_{-0.47}$ &   $0.50^{+0.25}_{-0.22}$ &   $0.62^{+0.26}_{-0.36}$ \\
     &      & ... Mg &           $0.95\pm0.10$ &  $3.40^{+0.32}_{-0.29}$ &   $1.23^{+0.07}_{-0.08}$ &   $1.16^{+0.07}_{-0.04}$ &            $1.13\pm0.09$ \\
     &      & ... Si &  $1.79^{+0.14}_{-0.16}$ &  $1.93^{+0.21}_{-0.20}$ &            $1.05\pm0.08$ &            $1.04\pm0.08$ &   $1.07^{+0.09}_{-0.08}$ \\
     &      & ... S &  $2.69^{+0.17}_{-0.16}$ &  $4.78^{+0.35}_{-0.33}$ &   $1.96^{+0.09}_{-0.11}$ &   $1.98^{+0.11}_{-0.10}$ &   $2.07^{+0.16}_{-0.14}$ \\
     &      & ... Ar &  $2.63^{+0.28}_{-0.26}$ &  $5.68^{+0.71}_{-0.67}$ &   $2.46^{+0.18}_{-0.23}$ &   $2.56^{+0.25}_{-0.24}$ &   $2.68^{+0.35}_{-0.32}$ \\
     &      & ... Ca &  $3.79^{+0.59}_{-0.58}$ &  $9.84^{+1.55}_{-1.45}$ &   $4.57^{+0.58}_{-0.55}$ &   $4.97^{+0.58}_{-0.62}$ &   $5.03^{+0.66}_{-0.77}$ \\
     &      & ... Fe $=$ Ni &  $2.57^{+0.21}_{-0.20}$ &  $6.41^{+0.60}_{-0.53}$ &   $0.83^{+0.14}_{-0.20}$ &   $1.04^{+0.14}_{-0.13}$ &   $1.10^{+0.31}_{-0.11}$ \\
     &      & ... Others & 1 (fixed) & 1 (fixed) & 1 (fixed) & 1 (fixed) & 1 (fixed) \\
     & $\log~\tau$ & (cm$^{-3}$ s) &          $10.83\pm0.03$ &          $10.48\pm0.01$ &           $11.17\pm0.07$ &           $11.21\pm0.05$ &  $11.26^{+0.06}_{-0.11}$ \\
     & $EM$ & ($10^{10}$ cm$^{-5}$) &    $40.0^{+4.1}_{-3.4}$ &    $10.3^{+1.0}_{-0.9}$ &     $86.6^{+8.9}_{-8.2}$ &     $97.7^{+9.4}_{-9.1}$ &   $94.8^{+15.4}_{-11.8}$ \\
Ejecta\,2 & $kT_{\rm e}$ & (keV) &            --- &            --- &   5 (fixed) &   5 (fixed) &            5 (fixed) \\
(c and e: \texttt{vvnei}, & $Z$ & ... Cr &                     --- &                     --- &   $53.5^{+81.6}_{-42.4}$ &    $54.9^{+62.8}_{-42.3}$ &   $50.9^{+52.0}_{-39.9}$ \\
~d: \texttt{vvpshock}) &      & ... Mn &                     --- &                     --- &  $73.6^{+130.0}_{-71.4}$ & $77.3^{+111.6}_{-76.5}$ &          $65.4^{<+82.2}$ \\
 &      & ... Fe $=$ Ni &                     --- &                     --- &    $18.0^{+11.5}_{-4.8}$ &  $14.3^{+6.5}_{-3.4}$ &    $12.9^{+11.4}_{-5.7}$ \\
     &      & ... Others & --- & --- & $=$ Ejecta\,1 & $=$ Ejecta\,1 & $=$ Ejecta\,1 \\
     & $\log~\tau$ & (cm$^{-3}$ s) &                     --- &                     --- &           $10.24\pm0.03$ & $10.53\pm0.04$\footnotemark[$*$] &  $10.34^{+0.18}_{-0.08}$ \\
     & $EM$ & ($10^{10}$ cm$^{-5}$) &                     --- &                     --- &   $1.93^{+0.66}_{-0.72}$ & $2.32^{+0.61}_{-0.69}$ &   $1.83^{+0.53}_{-0.66}$ \\
Ejecta\,3 & $kT_{\rm e}$ & (keV) &            --- &            --- &   --- &   --- &       $=$ Ejecta\,2 \\
(e: \texttt{vvnei})     & $Z$ &    &                 --- &                     --- &                      --- &                      --- &  $=$ Ejecta\,2 \\  
 & $\log~\tau$ & (cm$^{-3}$ s) &                     --- &                     --- &                      --- &                      --- &          $9.00^{<+0.56}$\footnotemark[$\dagger$] \\
 & $EM$ & ($10^{10}$ cm$^{-5}$) &                     --- &                     --- &                      --- &                      --- &   $0.74^{+0.96}_{-0.41}$ \\
\hline
$\chi^2$ & &  &        1410 &        1078 &          895 &          867 &          864 \\
d.o.f. & &  &        773 &        773 &          768 &          768 &          766 \\
\hline
\end{tabular}
\end{center}
\label{tab:parameters}
\begin{tabnote}
\hangindent6pt\noindent
\hbox to6pt{\footnotemark[$*$]\hss}\unskip%
The upper bound value of continuous $\tau$ distribution. The lower bound is set to $\log_{10}~\tau=8.0$.\\
\hbox to6pt{\footnotemark[$\dagger$]\hss}\unskip%
The lower end of statistical uncertainty was not found above the hard limit of the $\tau$ value at $\log_{10}~\tau=8.0$. \\
\end{tabnote}
\end{table*}%

\subsubsection{Multi-$\tau$ ejecta models}\label{sec:multitau}

To better describe the low ionization nature of the Fe~K$\alpha$ line in the full-band analysis, we introduced the second NEI component. In trial fittings, the electron temperature was inferred to be high, i.e. $>$3~keV. The temperature was as high as a few 10~keV. We found that the improvement of fits were small and only found at the Fe-L lines. Because the line centroid of Fe~K$\alpha$ does not change above $\approx$5~keV (figure~\ref{fig:contours}) and also the Fe-L structure has relatively large systematic uncertainty in emission models, we fixed the electron temperature of the second NEI at 5~keV.
Between the two ejecta components, the ionization timescale $\tau$, abundance $Z$ for Fe, and emission measure $EM$ were allowed to be different. Because the measured energy centroid ratios between the Cr~K$\alpha$, Mn~K$\alpha$, and Fe~K$\alpha$ lines are all consistent with being a single component (i.e., can be described by the same $kT_{\rm e}$ and $\tau$), Cr and Mn are likely associated with the Fe-dominated ejecta. Thus, the abundances of Cr and Mn were set free for the Fe-dominant ejecta while those for the IME-dominated ejecta were fixed to the solar values. The Ni abundance was assumed to be the same as Fe in relative values to the solar abundances because of no clear detection of the Ni~K$\alpha$ line. The abundances of the other elements were all tied to each other between the two ejecta components. This model (``c'') significantly improved the fit ($\chi^2/\rm{dof}=$1.17) as shown in figure~\ref{fig:specfit}c. The best-fit parameters are given in the column (c) of table~\ref{tab:parameters}. As already indicated by the band-limited spectral analysis, the ionization timescale $\tau$ of the Fe-dominated component (Ejecta\,2 in table~\ref{tab:parameters}) is about one-order-of-magnitude smaller than the IME-dominated one (Ejecta\,1). 

There are still small residuals in the Fe-K band with the two-ejecta model (figure~\ref{fig:specfit}c), particularly for the FI data. One is an asymmetric ``M''-shaped feature found around Fe~K$\alpha$ with larger excess at the lower-energy side. This indicates that the current model underestimates the width. Actually, the line required a larger width ($\approx60$~eV: section~\ref{sec:band-limited}) than that expected from the current model ($\approx 45$~eV). The excess broadening of $\approx$40~eV corresponds to a Doppler speed of $\sigma_v \approx $2000~km~s$^{-1}$ if bulk motions superposed on the line of sight are the cause. On the other hand, the He$\alpha$ lines of IME required no significant broadening ($\sigma<5$~eV at the 90\% upper limit). 
Because the equivalent-width maps by CO16 showed no clear indication of Fe ejecta localized outside IME, 
it is very unlikely that the Fe ejecta has a much larger bulk speed than the preceding IME ejecta, making the Doppler broadening origin unfeasible. Another possibility is thermal broadening. The 40 eV excess broadening converts to $kT_{\rm Fe}\approx$2~MeV, which is $\approx$400 times higher than $kT_{\rm e}$.

An alternative idea to explain the broadening is that the Fe ejecta consist of multiple components with different ionization timescales. This idea is supported by the possible Fe K$\beta$ line at 7.15~keV. The feature is not well reproduced by the current model. This is partly because the mismatch of the line centroid; Fe atoms in Ejecta\,2 are dominated by F- and O-like ions, whose Fe K$\beta$ line energies are expected around 7.35~keV. The insufficient line emissivity can also be attributed to the charge-state distribution in the current model. As ionization stage gets as high as O-like, Fe ions lose all $n=$3 electrons, which results in zero fluorescence yield for the K$\beta$ transition.  
Such a Fe-K structure is reminiscent of Tycho's remnant \citep{2014ApJ...780..136Y}. The Fe-K spectrum of Tycho's SNR is explained by a combination of two components with different ionization timescales. Therefore, we further examined models with multiple $\tau$ for the Fe-dominated ejecta. 

First we replaced Ejecta\,2 with a plane-parallel shock plasma by using \texttt{vvpshock} in Xspec, which assumes a continuous $\tau$ distribution with a constant differential emission measure against $\tau$. This model (``d'') mitigated the residuals at the Fe~K$\alpha$ and K$\beta$ lines with $\chi^2/\rm{dof}=$1.13 (figure~\ref{fig:specfit}d). We then tried another model that assumes two discrete $\tau$ values for the Fe-dominated ejecta by using two \texttt{vvnei} components (Ejecta\,2 for higher $\tau$ and Ejecta\,3 for lower $\tau$). All the parameters of Ejecta\,3 were fixed to those of Ejecta\,2 except for the emission measure ($EM$) and the ionization timescale ($\tau$). As shown in figure~\ref{fig:specfit}e, this model (``e'') better explained the Fe K$\beta$ feature than the previous one, although the improvement for the entire spectrum is marginal ($\approx$1$\sigma$ by F-test). The improvement from (c), on the other hand, is significant at $\approx$5$\sigma$. 
The best-fit parameters for the models are given in the columns (d) and (e) of table~\ref{tab:parameters}, respectively. 

Even with the model (d) or (e), we still see some line-like residuals. These are found at $\approx$1.55~keV and $\approx$2.60~keV. The former is near Al He$\alpha$ (1.598~keV for 1s--2p $^1$P$_1$ and 1.575~keV for 1s--2s $^1$S$_0$), but is slightly lower. The abundance of Al, when being set as a free parameter, was derived to be 1.62$^{+0.35}_{-0.29}$~solar. The super-solar abundance of Al is expected in ejecta of Type~Ia remnant, but both the line detection and measured abundance value of Al may require a future confirmation because of the discrepancy in the line energy. The latter feature at $\approx$2.60~keV indicates underestimation of S Ly$\alpha$/He$\alpha$ line ratio or the average charge of S ions, but we cannot conclude whether this is real or not because it is only visible in the FI spectrum.

\section{Discussion}

\subsection{Nature of Fe ejecta and chemical stratification}\label{sec:iron_ejecta}

As established in the spectral analysis, the Fe-dominated ejecta (Ejecta\,2 in table~\ref{tab:parameters}) has one-order-of-magnitude lower ionization timescale than the IME-dominated ejecta (Ejecta\,1). This is the evidence for the lag in the shock-heating ages between Fe and IME, which overturns the previous results of spectral analyses by CO16 and SE17. A plausible interpretation of the lag is a chemically stratified structure of the ejecta. The present data do not allow us to directly investigate a layered structure using X-ray images. CO16 argued that the Fe K emission is centrally concentrated compared to IME. This would support our argument, but no quantitative or statistical examination was made. A future study with a better spatial resolution and a large effective area up to the Fe-K band is needed.
The shock-heated Fe mass in Ejecta\,1 is about one-third of that in Ejecta\,2 (table~\ref{tab:snrparams}), 
indicating that a non-negligible fraction of the Fe ejecta exists in relatively outer regions of the remnant. The outer Fe ejecta may be formed either by an asymmetric explosion as suggested for some Ia SNRs (e.g., SN\,1006: \cite{2013ApJ...771...56U}) or by a partial mixing of the stratified structure. Note that, however, the Fe mass estimation of Ejecta\,1 is highly uncertain as it is primarily derived from the emission measure of Fe-L lines, and thus the Fe mass ratio may also be inaccurate. 

The spectral analysis also showed that an additional Fe-ejecta component with a lower ionization timescale may be needed (the models ``d'' or ``e'' in \S\ref{sec:multitau}). While the ionization structure in the model (d) is naturally expected even for a simple ejecta evolution \citep{2001ApJ...548..820B},  that in the model (e) requires some additional process. One possible cause is a large non-uniformity of the Fe ejecta; if a denser clump exists in ejecta (e.g., \cite{2001ApJ...549.1119W}), it gets larger $\tau$. Another possibility is anisotropy of reverse-shock arrival times due for instance to a gradient in the ambient density. Toward a denser direction, swept-up ISM mass dominates over ejecta mass more quickly, and forms a reverse shock earlier, which results in a higher value of the ionization timescale. Finally we address the third possibility, in which the dust contributes to X-ray emission. If a significant fraction of Fe is stored as dust, it is gradually destroyed and melts into hot gas, supplying quasi-neutral atoms in thermal plasma \citep{1997ApJ...477L..49B}. Such a component, if it exists, would have a lower ionization timescale compared to the bulk gas component. 

\begin{figure}[!htbp]
\begin{center}
\includegraphics[width=8cm]{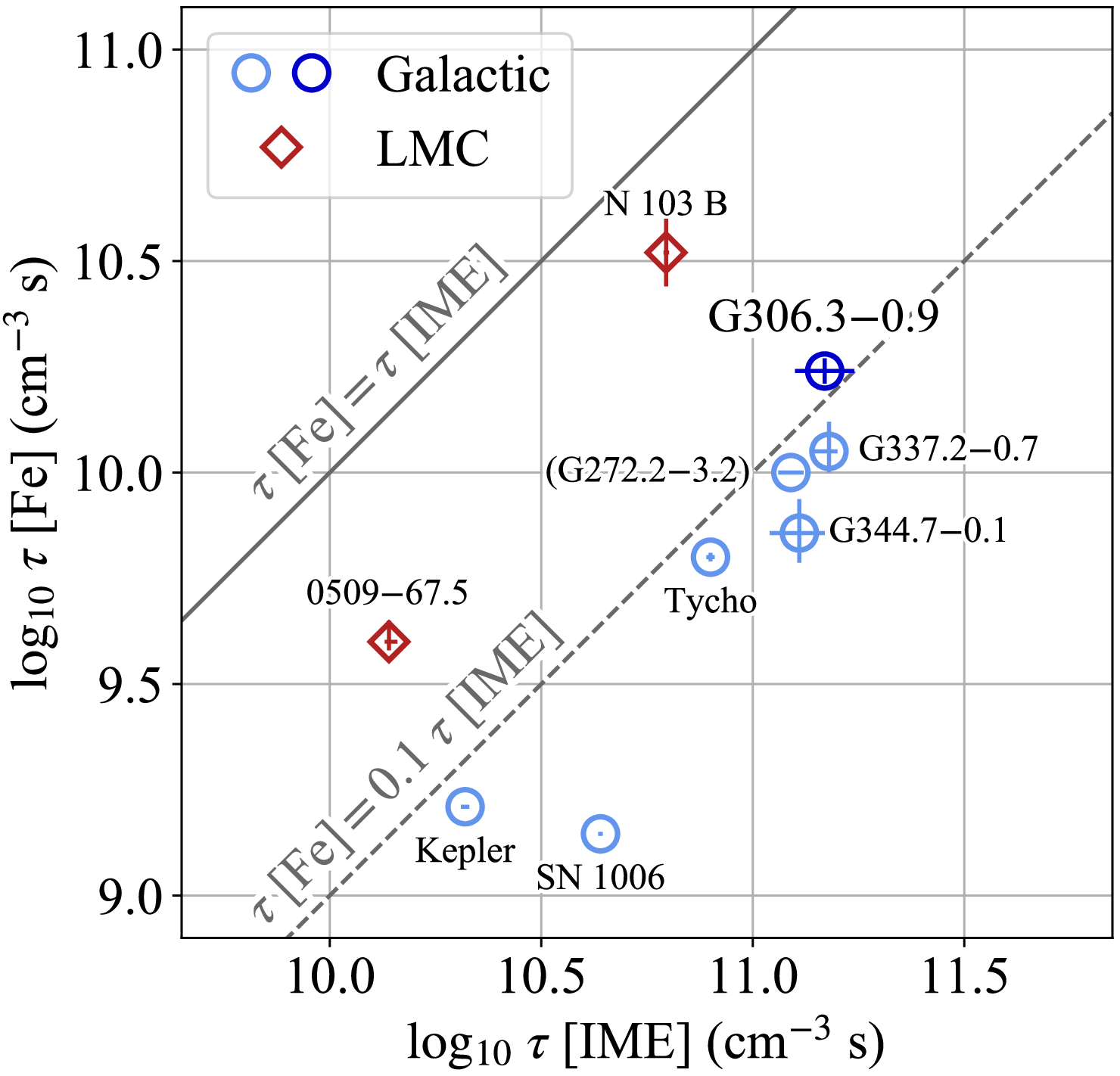}
\end{center}
\caption{Ionization timescales of Fe versus IME measured with Suzaku (Kepler's SNR, Tycho's SNR, and SNR\,0509$-$67.5: \cite{2015ApJ...808...49K}; SN\,1006: \cite{2013ApJ...771...56U}; G344.7$-$0.1: \cite{2012ApJ...749..137Y}; N\,103\,B: \cite{2014PASJ...66...26S}; G337.2$-$0.7: \cite{2016PASJ...68S...3T}; G306.3$-$0.9: this work). The light blue circles and red diamonds are the Galactic and LMC SNRs, respectively. Our result on G306.3$-$0.9 based on the model (c) is shown in blue. The $\tau$ value for the Fe ejecta in G272.2$-$3.2 \citep{2016PASJ...68S...7K} is an assumed one but also shown for reference. Error bars show 90\% confidence intervals.
}
\label{fig:tautau}
\end{figure}

Figure~\ref{fig:tautau} shows ionization timescales of IME-dominated and Fe-dominated ejecta for a compilation of Ia (or candidate) SNRs studied with Suzaku. Because the same instrument is used, this plot enables us a straightforward comparison of the properties of these remnants. We note that a systematic uncertainty originating from a possible variation in atomic codes is not shown here but likely not so large as it interferes with the discussion below (see appendix~\ref{app:atomic} for details). For those analyzed with the plane-parallel shock (\texttt{pshock}) model (G337.2$-$0.7: \cite{2016PASJ...68S...3T} for both IME and Fe; N\,103\,B: \cite{2014PASJ...66...26S} for Fe), only the upper-bound values $\tau_{\rm u}$ are available, and thus we estimate the average ionization timescales by fitting simulated \texttt{pshock} spectra with the simple \texttt{nei} model. We find that the approximated $\tau$ values are close to 0.5~$\tau_{\rm u}$ as naturally expected from the constant differential emission measure distribution in \texttt{pshock}.
The Galactic Ia SNRs show a systematic trend of distribution described as $\tau$ for Fe $\approx$~0.1~$\tau$ for IME.
G306.3$-$0.9 slightly extends this relation and occupies the top-right end, indicating that it is one of the most thermally evolved systems. 
The plot illustrates that the lagged shock heating of Fe ejecta is ubiquitous in Galactic Ia SNRs over one order of magnitude of $\tau$. 
The trend could be explained at least partially by stratification of ejecta in Type~Ia SNRs. If these SNRs indeed have stratified ejecta, our results suggest that the mixing of ejecta is not so efficient even in a relatively later stage of the Sedov phase, where the interaction with ambient materials is not negligible. 
However, physical reality is undoubtedly more complex and we need observations with higher angular resolution and high spectral resolution to understand this better.

\subsection{Updating the distance}\label{sec:distance}

\begin{figure*}[htbp]
\begin{center}
\includegraphics[width=17cm]{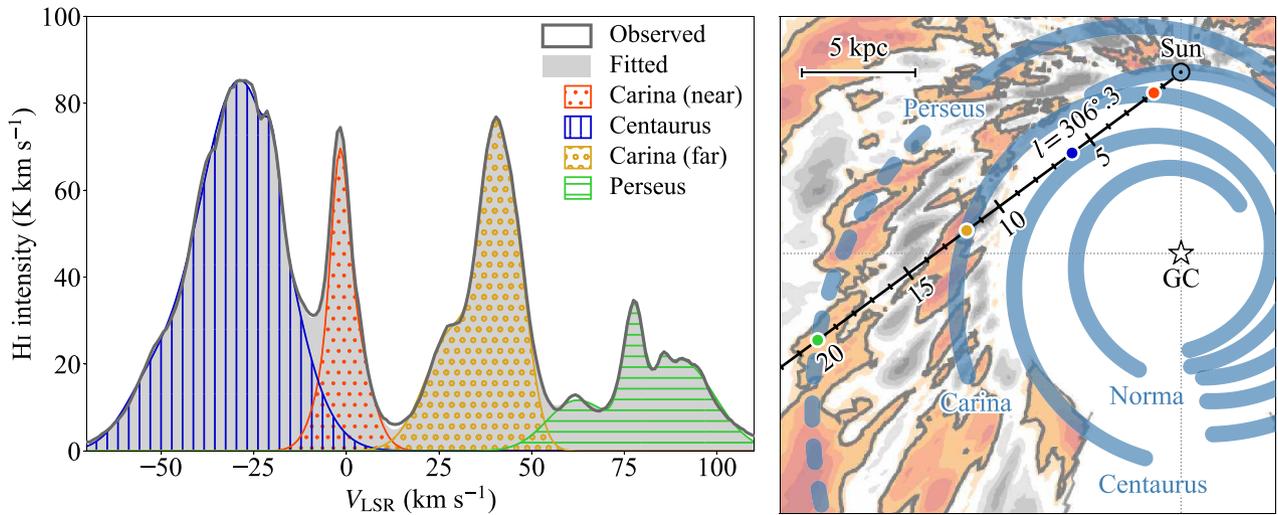}
\end{center}
\caption{Left: The H\emissiontype{I} velocity profile with the GASS (thick grey curve) fitted by an empirical model (light grey area; see text for details). Individual contributions from the four regions at different distances (table~\ref{tab:arms}) are shown with colored hatches. 
Right: A face-on view of the Galactic structures. The solid blue curves show the spiral arms of the inner Galaxy and the vicinity of the Sun based on the polynomial-logarithmic model \citep{2014A&A...569A.125H}. The contours are the H\emissiontype{I} surface density perturbation map \citep{2006Sci...312.1773L} tracing the spiral arms in the outer Galaxy (orange areas). The maps are scaled assuming the distance to the Galactic center of 8~kpc. The dashed blue curve is an eye-guide for the outer extension of the Perseus arm. The major structures along the line of sight toward G306.3$-$0.9 (black line) are marked with the dots with the matched colors with the left panel. 
}
\label{fig:gassv}
\end{figure*}

\begin{table*}[htdp]
\caption{The relative contributions from individual Galactic arms to the absorption column in the direction of G306.3$-$0.9.}
\label{tab:arms}
\begin{center}
\begin{tabular}{lcccccccc}
\hline
          Arm & $d$ (kpc) & $I_{\rm H\emissiontype{I}}$ & $N_{\rm H\emissiontype{I}}$ & $f_{\rm mol}$ & $N_{\rm H_2}$ & $N_{\rm H\emissiontype{I}+H_2}$ \\
\hline
Carina (near) &             $\sim$1 &                         6.8 &                        0.12 &          0.24 &          0.04 &                            0.16 \\
Centaurus     &             $\sim$6 &                        27.3 &                        0.50 &          0.36 &          0.29 &                            0.78 \\
Carina (far)  &            $\sim$12 &                        14.2 &                        0.26 &          0.04 &          0.01 &                            0.27 \\
Perseus       &            $\sim$20 &                         8.8 &                        0.16 &          0.00 &          0.00 &                            0.16 \\
\hline
Total         &                 --- &                        57.1 &                        1.04 &           --- &          0.34 &                            1.38 \\
\hline
\end{tabular}
\end{center}
\begin{tabnote}
\hangindent6pt\noindent
\hbox to6pt{\footnotemark[$*$]\hss}\unskip%
Relative H\emissiontype{I} intensity derived from the GASS profile. \\
\hbox to6pt{\footnotemark[$\dagger$]\hss}\unskip%
Individual contribution from each structure in column density in units of $10^{22}$~H~cm$^{-2}$. The line-of-sight distribution of the molecular component is assumed to be the same as the atomic component. \\
\end{tabnote}
\end{table*}

In the previous X-ray studies (\cite{2013ApJ...766..112R}, CO16, and SE17), the distance has been assumed to be 8~kpc, a fiducial value for the Galactic center (e.g., \cite{1993ARA&A..31..345R}; \cite{2018PASP..130b4101C}). This is consistent with the fact that we do not find H$\alpha$ emission associated to the remnant in the SuperCOSMOS images \citep{2005MNRAS.362..689P}, setting a lower limit of the distance to be 4~kpc \citep{1988A&A...205...95G}. 
The reported X-ray absorption columns, $N_{\rm H}=($1.5--2.0$)\times 10^{22}$~H~cm$^{-2}$, are indeed within a typical range for objects in the Galactic center region. However, the average ISM density in the direction of G306.3$-$0.9 is likely much lower, suggesting a farther location of the remnant than 8~kpc. This is supported by the fact that the reported  $N_{\rm H}$ values are comparable to the Galactic total hydrogen column of 1.4$\times 10^{22}$~H~cm$^{-2}$ along the line of sight (see below). Thus, we further examine possible location of the remnant using the velocity-decomposed Galactic hydrogen distribution.

The left panel of figure~\ref{fig:gassv} shows the H\emissiontype{I} 21~cm emission velocity profile toward G306.3$-$0.9 with the Parkes Galactic All Sky Survey (GASS: \cite{2015A&A...578A..78K}). There are four major velocity groups arising from spiral arms at different distances $d$ (table~\ref{tab:arms}): the Centaurus arm at $V_{\rm LSR} \approx -$30~km~s$^{-1}$, the near and far sides of the Carina arm at $-$2 and $+$40~km~s$^{-1}$, respectively, and the outer extension of the Perseus arm around $+$80~km~s$^{-1}$, where $V_{\rm LSR}$ is the velocity at the local standard of rest. The locations of these intervening structures are shown in the right panel of figure~\ref{fig:gassv}. The column densities contributed by the atomic component ($N_{\rm H\emissiontype{I}}$) can be directly estimated from the H\emissiontype{I} intensity. We fit the profile with an empirical model consisting of 12 Gaussians, and derive $N_{\rm H\emissiontype{I}}$ for each arm by multiplying a conversion factor of 1.82$\times$10$^{18}$~H~cm$^{-2}$~(km~s$^{-1}$)$^{-1}$ \citep{1978ppim.book.....S}. For the molecular component ($N_{\rm H_2}$), we utilize the molecular fraction, $f_{\rm mol} \equiv \Sigma_{\rm H_2} / (\Sigma_{\rm H\emissiontype{I}} + \Sigma_{\rm H_2})$, where $\Sigma$ is the surface number density of hydrogen atoms \citep{2016PASJ...68....5N}, as an indicator of the column density ratio along the line of sight, i.e., $N_{\rm H_2} \approx f_{\rm mol} / (1 - f_{\rm mol}) \cdot N_{\rm H\emissiontype{I}}$. 
The total hydrogen column densities for the individual arms are given in table~\ref{tab:arms}. 

The $N_{\rm H}$ values derived in the spectral fittings correspond to 88--92\% (models c--e; table~\ref{tab:parameters}) of the Galactic total. 
This indicates that the most likely location of G306.3$-$0.9 is in the outer extension of the Perseus arm at $\sim$20~kpc, which corresponds to the column density of $\ge$88\% of the Galactic total. This is significantly farther than the previously assumed distance, but it is reasonable because in the direction to the remnant the average density is lower than that toward the Galactic center. 
We note that our estimation has systematic uncertainty. The ignorance of the contribution from ionized material in table~\ref{tab:arms} would underestimate the total Galactic hydrogen column. Thus the distance might be as near as the far side of the Carina arm ($\sim$12~kpc), but the 8~kpc distance is likely ruled out because the intervening structures in this case (the near Carina and Centaurus arms) account for only $\approx$70\% of the Galactic total, leaving a significantly large deviation from the measured $N_{\rm H}$ value. The 8~kpc distance would be even less likely if we adopt an alternative X-ray $N_{\rm H}$ value using an absorption model with a newer photo-electric cross section (\texttt{tbabs}: \cite{2000ApJ...542..914W}); we get a 50\% larger $N_{\rm H}$, preferring an even farther distance to the remnant. The different systematic factors may partially cancel to each other. We hereafter adopt 20~kpc as the distance to this remnant.

\subsection{Global properties and evolutionary stage}\label{sec:global}

With the parameters obtained by the spectral analysis (section~\ref{sec:spectra}) and the updated distance (section~\ref{sec:distance}), we here estimate global properties of the SNR G306.3$-$0.9. At 20~kpc, the angular radius of \timeform{110''} converts to $R=$11~pc. We derive the volume emission measure ($VEM_{\rm X}$), ion density ($n_{\rm X}$), and mass ($M_{\rm X}$) for each thermal emission component, as presented in table~\ref{tab:snrparams}, based on the best-fit parameters with the multi-$\tau$ model (c). Here we assume the total emitting region to be a hemisphere with 11~pc radius because only the southern hemisphere is bright in the Chandra images (CO16). For the ejecta, we derive parameters individually for heavy elements with line detections. Volume filling factors are not taken into account, and thus the densities should be considered as lower limits and the masses are upper limits. 
These parameters scale as $d^{2}$, $d^{-0.5}$, and $d^{2.5}$, respectively, where $d$ is the distance to the remnant. 

\begin{table}[!htdp]
\caption{Properties of ISM and ejecta based on the model (c).}
\begin{center}
\begin{tabular}{lccc}
\hline
{} &                      $VEM_{\rm X}$\footnotemark[$*$]  &                      $n_{\rm X}$\footnotemark[$\dagger$] &    $M_{\rm X}$\footnotemark[$\ddagger$]  \\
& ($10^{59}$~cm$^{-3}$) & (cm$^{-3}$) & ($M_{\solar}$) \\
\hline
\multicolumn{4}{c}{\dotfill ISM\dotfill}\\
         &   $1.4\pm0.7$ &                         $0.88^{+0.20}_{-0.25}$ &  $(1.5\pm0.4)\times$10$^2$ \\
\hline
{} &                      $VEM_{\rm X}$\footnotemark[$*$]  &                      $n_{\rm X}$\footnotemark[$\dagger$] &    $M_{\rm X}$\footnotemark[$\ddagger$]  \\
& ($10^{54}$~cm$^{-3}$) & (10$^{-5}$ cm$^{-3}$) & (10$^{-2}$ $M_{\solar}$) \\
\hline
\multicolumn{4}{c}{\dotfill Ejecta\,1\dotfill} \\
Ne &                $3.2^{+1.5}_{-2.4}$ & $3.7^{+1.8}_{-2.8}$ & $9.2^{+4.4}_{-7.0}$ \\
Mg &                        $1.9\pm0.2$ &         $2.3\pm0.2$ &         $6.8\pm0.6$ \\
Si &                        $1.5\pm0.2$ &         $1.8\pm0.2$ &         $6.2\pm0.6$ \\
S  &                        $1.3\pm0.1$ &         $1.5\pm0.1$ &         $6.1\pm0.5$ \\
Ar &                      $0.37\pm0.05$ & $0.43^{+0.04}_{-0.05}$ &         $2.1\pm0.2$ \\
Ca &                      $0.43\pm0.07$ &         $0.50\pm0.07$ & $2.5^{+0.4}_{-0.3}$ \\
Fe &                $1.6^{+0.3}_{-0.4}$ & $1.9^{+0.3}_{-0.5}$ &             $13^{+2}_{-3}$ \\
\multicolumn{4}{c}{\dotfill Ejecta\,2\dotfill} \\
Cr & $0.023^{+0.036}_{-0.020}$ & $0.18^{+0.28}_{-0.15}$ & $1.2^{+1.8}_{-1.0}$ \\
Mn &       $0.017^{<+0.030}$ &       $0.13^{<+0.23}$ &      $0.9^{<+1.6}$ \\
Fe &             $0.77^{+0.56}_{-0.36}$ & $6.0^{+4.1}_{-2.2}$ &             $42^{+29}_{-16}$ \\
\hline
\end{tabular}
\end{center}
\label{tab:snrparams}
\begin{tabnote}
\hangindent6pt\noindent
\hbox to6pt{\footnotemark[$*$]\hss}\unskip%
The volume emission measure of H for ISM or that of each element for ejecta. \\
\hbox to6pt{\footnotemark[$\dagger$]\hss}\unskip%
The ion density of H for ISM or that of each element for ejecta derived from $VEM_{\rm X}=n_{\rm e}n_{\rm X}V$, $n_{\rm e} \approx 1.2n_{\rm H}$, and $n_{\rm X} = Z_{\rm X} n_{\rm H}$, where $Z_{\rm X}$ is the abundance of the element X in the number ratio to H. \\
\hbox to6pt{\footnotemark[$\ddagger$]\hss}\unskip%
The mass of each component or element, $M_{\rm X} = m_{\rm X} n_{\rm X} V$. For ISM, the total mass including all the elements with the solar abundances is calculated, i.e., $m_{\rm X} = \mu m_{\rm H}$, where $\mu \approx 1.4$ is the mean atomic mass relative to H. 
\end{tabnote}
\end{table}

The total mass of the shocked ejecta is $\approx$1~$M_{\solar}$ at 20~kpc. This is in a good agreement with a view that the whole Ia ejecta has been shock heated. The swept-up mass of the shocked ISM ($\approx$150~$M_{\solar}$) is much greater than the ejecta mass. 
The Sedov timescale $t_{\rm s}$, blast-wave shock speed $v_{\rm s}=0.4R/t_{\rm s}$, and post-shock temperature $T_{\rm s}$ are estimated as follows:
\begin{eqnarray}
t_{\rm s} &=& 6~\left(\frac{E_{51}}{d_{20}^5~\rho_1}\right)^{-1/2}~\rm{kyr}, \\
v_{\rm s} &=& 700~\left(\frac{E_{51}}{d_{20}^3~\rho_1}\right)^{1/2}~\rm{km~s}^{-1}, \\
kT_{\rm s} &=& 0.6~\left(\frac{E_{51}}{d_{20}^3~\rho_1}\right)~\rm{keV},
\end{eqnarray}
where $d_{20}$, $\rho_1$, and $E_{51}$ are the distance, ambient medium mass density, and explosion energy, in units of 20~kpc, 1~amu~cm$^{-3}$, and $10^{51}$~erg, respectively. As shown in the equations, we leave the ambient mass density as a parameter, $\rho_1$, rather than estimate it from the X-ray spectral fits. This is mainly because it would largely be affected by systematic factors like the filling factor. 
The series of evidence consistently supports that the SNR has already been entered into the Sedov phase. The thermal timescale of ejecta after shock heating can be evaluated from the ionization timescale, which is $t_{\rm ioni}\approx$4.7$~(n_{\rm e}/1~{\rm cm}^{-3})^{-1}$~kyr for the IME-dominated ejecta. This is consistent with the age of the remnant, whose upper limit is set by the Sedov timescale derived above. 

One may consider that the electron temperature of the Fe ejecta (5 keV) is unusually high for the ejecta with an SNR age of $\sim$6~kyr. Similarly high $kT_{\rm e}$ values have been observed in some SNRs, e.g., G272.2$-$3.2 (2.8~keV; \cite{2016PASJ...68S...7K}) and G337.2$-$0.7 (3.1~keV; \cite{2016PASJ...68S...3T}). These SNRs also have similar ionization timescales for IME (figure~\ref{fig:tautau}) or a dynamical timescale ($\sim$6~kyr for G272.2$-$3.2). The highest temperature components are always associated with the Fe ejecta. This can naturally be explained by the fact that the reverse-shock speed is initially low in frame of unshocked ejecta but gradually gets higher as the shock penetrates into the inner region \citep{1999ApJS..120..299T}, where the Fe-ejecta concentration is expected if stratified. 

In the spectral analysis, we found the excesses in the broadening at Fe K$\alpha$ ($\approx$40~eV) and in the intensity of Fe K$\beta$ with the single-$\tau$ Fe ejecta modeling (the model c), which were later explained by the multi-$\tau$ components (in the model d or e). Here, we briefly revisit other possibilities for the origin of the broadening in terms of consistency with the our findings discussed above.
The broadening corresponds to the velocity dispersion of $\sigma_v\approx$2000~km~s$^{-1}$ if it is entirely due to the bulk motion. This is comparable to the Sedov velocity ($v_{\rm s}=0.4R/t_{\rm s}\sim$700~km~s$^{-1}$). Alternatively, in the case of thermal broadening,  the post-shock Fe ion temperature of 2~MeV converts to the reverse-shock speed of $\sim$5500~km~s$^{-1}$ if Coulomb cooling is ignored. These estimates on the expansion speed ($\sim$700~km~s$^{-1}$) and reverse shock speed ($\sim$5500~km~s$^{-1}$) consistently predict the shock-heated age of Fe at $\sim$90\% of the age at which the reverse shock reaches to the center of the ejecta, using the analytical model by \citet{1999ApJS..120..299T}. The radial location of shock heated Fe is estimated to be at $\sim$20\% relative to the outer boundary of the ejecta (or contact discontinuity; \cite{2017MNRAS.465.3793T}). These numbers may have considerable uncertainties, but qualitatively agree with the global properties, namely that the SNR has entered into the late Sedov phase and the reverse shock has been reached to the innermost part where Fe is located.

The Ia origin argued by the previous studies is confirmed with our updated ejecta modeling. The Fe-K progenitor typing method \citep{2014ApJ...785L..27Y} can discriminate Ia and core-collapse origins and is even possible to constrain subtypes of Ia SNRs by combining line centroid and intensity of the Fe~K$\alpha$ line. 
The unabsorbed flux of the Fe~K$\alpha$ line was measured at $(9.91^{+0.76}_{-0.77})\times10^{-6}$~photons~cm$^{-2}$~s$^{-1}$. This corresponds to the luminosity of $4.7\times10^{41}$~photons~s$^{-1}$ at the 20 kpc distance. Combined with the line centroid at $\approx6.48$~keV, this puts G306.3$-$0.9 at the middle of the allowed region for Ia SNRs (figure~1 of \cite{2014ApJ...785L..27Y}). A recent proposal to distinguish Ia and core-collapse SNRs using the Fe/Ne ejecta mass ratio \citep{2016PASJ...68S...9T} can also be applied. In this method, Fe ejecta is expected to have a larger mass than Ne in the case of Ia SNRs. From table~\ref{tab:snrparams}, we estimate the ratio to be 4.5$^{+4.6}_{-2.7}$, again confirming the Ia origin. The relative abundances of trace elements to Fe are also the key to constraining explosion mechanisms and progenitor subtypes. The Cr-to-Fe abundance ratio is one example \citep{2013ApJ...771...56U}. With the current data, however, the ratio $Z_{\rm Cr}/Z_{\rm Fe}$ is poorly constrained to be 3.0$^{<+4.6}$~solar, which prevents us from distinguishing different explosion mechanisms between delayed detonation model and classical deflagration model \citep{1999ApJS..125..439I}. 

\subsection{Future prospects}

\begin{figure}[!htbp]
\begin{center}
\includegraphics[width=8cm]{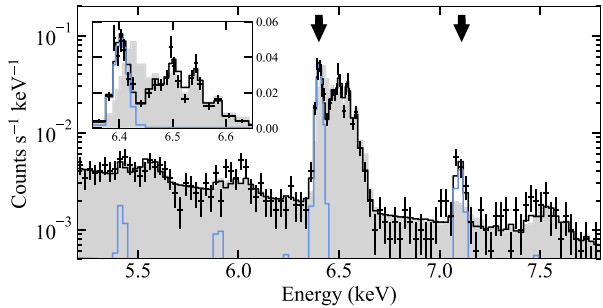}
\end{center}
\caption{A 200 ks simulation spectrum of the Fe-K band of G306.3$-$0.9 for Resolve on XRISM (the same response as the SXS on Hitomi is assumed). The black crosses are the simulated data based on the model (e) shown with the black histogram. The light blue histogram indicates the contribution from the lower $\tau$ component of the Fe-dominated ejecta (Ejecta\,3 in table~\ref{tab:parameters}). For comparison, the model (d) is shown with the gray filled histogram. The Gaussian broadening with a velocity dispersion of $\sigma_{v} = $500~km~s$^{-1}$ is assumed based on the forward-shock velocity derived in section~\ref{sec:global}. The arrows indicate the structures with large differences between the two models.}
\label{fig:sxssim}
\end{figure}

High-resolution spectroscopy with micro-calorimeter spectrometers to be flown on X-ray Imaging and Spectroscopy Mission (XRISM, formerly XARM) and Athena \citep{2018arXiv180706903G} will extend the spectroscopic study of ejecta ionization structure in Type~Ia SNRs. The energy resolution of $\le $5~eV combined with the energy calibration accuracy of $\le $1--2~eV makes the ionization timescale measurements much more accurate. Also, as demonstrated by the recent Hitomi observation of the SNR N\,132\,D \citep{2018PASJ...70...16H}, the $\approx $30 times better energy resolution enables us to detect and characterize Fe~K$\alpha$ even with limited photon counts, extending the sample toward fainter SNRs. Also expected is to better understand detailed nature of Fe ejecta, for instance multi-$\tau$ components as suggested by the Suzaku spectrum in this work (section~\ref{sec:iron_ejecta}). Here we briefly demonstrate an example simulation targeted for G306.3$-$0.9. 

Figure~\ref{fig:sxssim} is a simulated spectrum for the Resolve micro-calorimeter spectrometer on XRISM. With a reasonable exposure time, we can easily distinguish the two competing models with low ionization Fe. If the Fe ejecta has a discrete $\tau$ distribution (the model e, the black crosses and histogram), then the spectrum exhibits narrow peaks at 6.40 and 7.10~keV (the black arrows). These features have significantly different energies or intensities in the case of the continuous $\tau$ under the plane-parallel shock condition (the model d, the gray filled histogram), where the fraction of low-ionization ions of $\log_{10} \tau \lesssim $9 is considerably suppressed. We note that the detection of Ni~K$\alpha$ at $\approx $7.5~keV is also expected. 

Finally we note that, in order to constrain the detailed ionization structure in SNRs, not only high-resolution spectra but also accurate and complete emission modeling is necessary. Recently Hitomi observations of the Perseus cluster benchmarked atomic codes for CIE plasma \citep{2018PASJ...70...12H}. There is no such systematic benchmarks on NEI plasma, but discrepancies between the existing atomic codes (appendix~\ref{app:atomic}) clearly indicate the need of substantial updates of the atomic codes with laboratory data obtained by e.g., electron beam ion traps.

\section{Summary}

We have presented a new Suzaku analysis result of the Galactic SNR G306.3$-$0.9. The absolute energy scale of the XIS at 6--7 keV was cross-calibrated to the micro-calorimeter data using the Perseus cluster observations to study detailed thermal nature of the Fe ejecta. We discover the lower ionization timescale of Fe than IME using the band-limited fittings, and establish multi-component ejecta models to describe the full band spectrum. The low ionization of Fe strongly suggests the chemically stratified structure of the ejecta, overturning the previous results. We refine the distance to the remnant to be $\sim$20~kpc using the H\emissiontype{I} velocity profile and X-ray absorption column density. By combining the large ISM-to-ejecta mass ratio of $\sim$100, dynamical timescale of $\sim$6~kyr, and lower ionization timescale for the Fe ejecta, we identify a unique situation of this SNR where a possibly stratified ejecta survive in a dynamically evolved system. 

\vspace{1cm}

The authors thank the referees for the fruitful comments. We appreciate the supports from the Suzaku and Hitomi teams. We are grateful to Shinya Nakashima and Hiroya Yamaguchi for the inputs on the XIS calibration and discussion, respectively. We also thank Jelle Kaastra and Adam R. Foster for checking the appendix. This work was supported in part by Grant-in-Aid for Scientific Research of the Ministry of Education, Culture, Sports, Science and Technology (MEXT) of Japan, No. 15K17657 (MS), 26800102 (HU), 18J01417 (HM), 15K05107 (AB), 15H02090 (TGT), and 25109004 (TGT and TT). 

\clearpage

\begin{appendix}

\section{Fe~K$\alpha$ centroids in ionizing plasma with various atomic codes}\label{app:atomic}

\begin{figure*}[!htbp]
\begin{center}
\includegraphics[width=17cm]{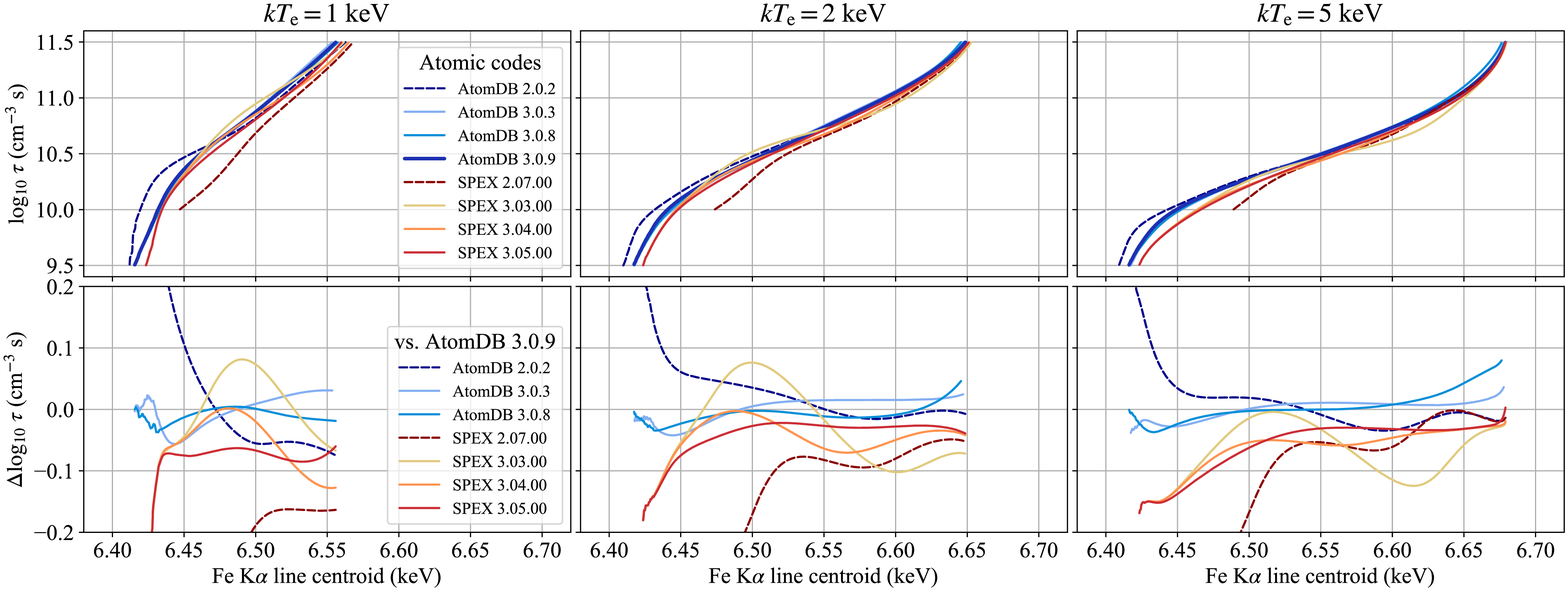}
\end{center}
\caption{Ionization timescales based on the Fe~K$\alpha$ centroids with various atomic codes (top) and their differences from AtomDB version 3.0.9 (bottom) for $kT_{\rm e} = $1 (left), 2 (center), and 5~keV (right). The compared atomic codes are AtomDB (cool colors) versions 2.0.2, 3.0.3, 3.0.8, and 3.0.9 and SPEX (warm colors) versions 2.07, 3.03, 3.04, and 3.05. The recent and old generations of the codes are shown in solid and dashed curves, respectively.}
\label{fig:atomic_tau}
\end{figure*}

\begin{figure*}[!htbp]
\begin{center}
\includegraphics[width=17cm]{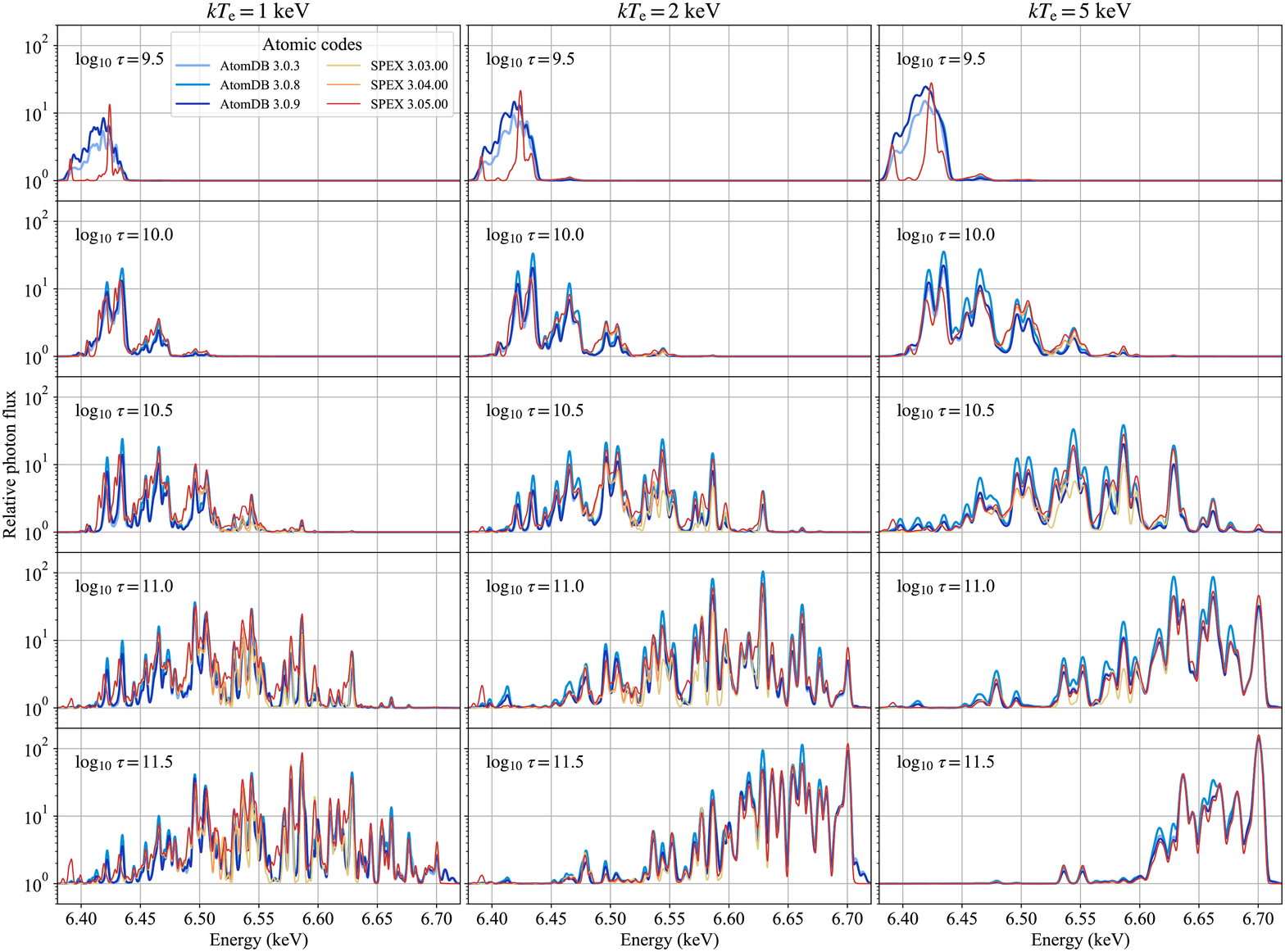}
\end{center}
\caption{The Fe~K$\alpha$ spectra for the recent versions of the atomic codes for $kT_{\rm e} = $1 (left), 2 (center), and 5~keV (right). The vertical axes show photon fluxes normalized by those of the underlying continuum (mainly bremsstrahlung). For each $kT_{\rm e}$, the evolution of the spectra is shown from top to bottom as a function of $\tau$. The color code is the same as in figure~\ref{fig:atomic_tau}.}
\label{fig:atomic_spec}
\end{figure*}

Because the measurement of $\tau$ relies on the atomic databases and spectral models (hereafter referred to as atomic codes), here we summarize comparisons of calculations with different atomic codes. Figure~\ref{fig:atomic_tau} shows ionization timescales based on the Fe~K$\alpha$ centroids with major atomic codes, AtomDB/APEC and SPEX. Model spectra were convolved with the XIS response and fitted in 6.0--6.9~keV with an empirical model of a Gaussian plus bremsstrahlung. For the SPEX models, we imported the simulated spectra into Xspec using \texttt{flx2tab} and the \texttt{atable} model and analyzed them exactly in the same manner as the AtomDB models to eliminate systematic differences in the fitting algorithms.
In the old generation of the codes (version 2), the large inter-code deviations of $\approx$50~eV in the Fe~K$\alpha$ line centroid or 0.5 in $\log_{10} \tau$ are seen. For the recent codes (version 3), the deviations are greatly reduced to the level within $\pm$0.1 in $\log_{10} \tau$ except for the very low ionization timescales of $\log_{10}~\tau < $10.0, but are still substantial even for the CCD resolution. 
These give a measure of typical systematic uncertainty in observational constraints on $\tau$ (e.g., in figure~\ref{fig:tautau}). 
We note that the deviations between different versions of the same code (e.g., SPEX versions 3.04 and 3.05) can be as large as those between the different codes (AtomDB and SPEX) depending on the parameter range. Therefore it is important for both reproducibility and clarifying a potential systematic uncertainty to describe in a paper which version of an atomic code is used in the spectral analysis, as done for analysis software packages and calibration databases. 

To get a basic idea on which factors are essential for the inter-code differences, we further compare the Fe~K$\alpha$ spectra in figure~\ref{fig:atomic_spec} for selected $\tau$ values with the recent versions of the atomic codes. Thermal broadening assuming electron-ion temperature equilibrium is applied\footnote{For AtomDB/APEC, the \texttt{APECTHERMAL} option in Xspec does not apply thermal broadening to relatively weak lines. Here, we instead applied a Gaussian smoothing (\texttt{gsmooth}) to the entire spectra. We note that it is also possible to change the minimum flux for broadening the lines by setting the \texttt{APECMINFLUX} parameter. See $\langle$https://heasarc.gsfc.nasa.gov/xanadu/xspec/manual/node134.html$\rangle$.} but no convolution is made for an instrumental response.
The immediate result from the comparison is that the number of emission lines and energies of individual lines are the major factor on the deviations of $\tau$ for the low ionization timescale of $\log_{10}~\tau\approx $9.5. This is not surprising as the atomic data are usually updated from higher to lower ionization sequences (for instance, in SPEX, the atomic line list of Fe ions for Mg-like or lower has not yet been updated from version 2). On the other hand, this is not the case for the higher $\tau$ values: the differences are rather seen in the relative intensities of individual lines. The major source of the deviations for $\log_{10} \tau\ge $10 is therefore either those in charge-state distribution (\cite{2009ApJ...691.1540B} for AtomDB and \cite{2017A&A...601A..85U} for SPEX) or in relative emissivities particularly of inner-shell excitation and ionization lines. 
For instance, the charge-state distribution calculation by \citet{2009ApJ...691.1540B} does not take into account the multiple-electron ionization due to inner-shell ionization followed by one or more Auger ejections, which partly accounts for the positive offset in $\tau$ of AtomDB 3.0.9 compared to SPEX 3.05. 
Even for relatively well calibrated emission lines such as the He-like triplets and the satellite lines of Li-like through C-like Fe ions in 6.55--6.70~keV, we see many code-to-code and version-to-version differences, but for the latest codes (AtomDB 3.0.9 and SPEX 3.05), these lines show a close agreement in both the line energies and emissivities. This is likely due to the recent updates in the calculations of inner-shell processes: a bug fix on a factor-of-two overestimated fluorescence emissivities for Li-like through F-like Fe ions in AtomDB (from 3.0.8 to 3.0.9)\footnote{See $\langle$http://www.atomdb.org/download.php$\rangle$.} and improved Auger rates for Be-like through N-like Fe ions in SPEX (from 3.03 based on \cite{1993A&AS...97..443K} to 3.05)\footnote{See $\langle$https://www.sron.nl/astrophysics-spex/manual$\rangle$.}. 
The agreement, however, degrades toward the lower ionization lines, which results in a larger systematic uncertainty in the $\tau$ measurement (figure~\ref{fig:atomic_tau}). This fact clearly demonstrates the need for further improvements on the atomic database and modeling particularly for NEI spectra. 

\end{appendix}

\bibliographystyle{aa}
\bibliography{g306_pasj}

\begin{thebibliography}{66}
\expandafter\ifx\csname natexlab\endcsname\relax\def\natexlab#1{#1}\fi

\bibitem[{{Anders} \& {Grevesse}(1989)}]{1989GeCoA..53..197A}
{Anders}, E. \& {Grevesse}, N. 1989, \gca, 53, 197

\bibitem[{{Badenes} {et~al.}(2005){Badenes}, {Borkowski}, \&
  {Bravo}}]{2005ApJ...624..198B}
{Badenes}, C., {Borkowski}, K.~J., \& {Bravo}, E. 2005, \apj, 624, 198

\bibitem[{{Badenes} {et~al.}(2007){Badenes}, {Hughes}, {Bravo}, \&
  {Langer}}]{2007ApJ...662..472B}
{Badenes}, C., {Hughes}, J.~P., {Bravo}, E., \& {Langer}, N. 2007, \apj, 662,
  472

\bibitem[{{Balucinska-Church} \& {McCammon}(1992)}]{1992ApJ...400..699B}
{Balucinska-Church}, M. \& {McCammon}, D. 1992, \apj, 400, 699

\bibitem[{{Borkowski} {et~al.}(2001){Borkowski}, {Lyerly}, \&
  {Reynolds}}]{2001ApJ...548..820B}
{Borkowski}, K.~J., {Lyerly}, W.~J., \& {Reynolds}, S.~P. 2001, \apj, 548, 820

\bibitem[{{Borkowski} \& {Szymkowiak}(1997)}]{1997ApJ...477L..49B}
{Borkowski}, K.~J. \& {Szymkowiak}, A.~E. 1997, \apjl, 477, L49

\bibitem[{{Brickhouse} {et~al.}(2000){Brickhouse}, {Dupree}, {Edgar},
  {Liedahl}, {Drake}, {White}, \& {Singh}}]{2000ApJ...530..387B}
{Brickhouse}, N.~S., {Dupree}, A.~K., {Edgar}, R.~J., {et~al.} 2000, \apj, 530,
  387

\bibitem[{{Bryans} {et~al.}(2009){Bryans}, {Landi}, \&
  {Savin}}]{2009ApJ...691.1540B}
{Bryans}, P., {Landi}, E., \& {Savin}, D.~W. 2009, \apj, 691, 1540

\bibitem[{{Camarillo} {et~al.}(2018){Camarillo}, {Mathur}, {Mitchell}, \&
  {Ratra}}]{2018PASP..130b4101C}
{Camarillo}, T., {Mathur}, V., {Mitchell}, T., \& {Ratra}, B. 2018, \pasp, 130,
  024101

\bibitem[{{Chevalier} {et~al.}(1992){Chevalier}, {Blondin}, \&
  {Emmering}}]{1992ApJ...392..118C}
{Chevalier}, R.~A., {Blondin}, J.~M., \& {Emmering}, R.~T. 1992, \apj, 392, 118

\bibitem[{{Combi} {et~al.}(2016){Combi}, {Garc{\'{\i}}a}, {Su{\'a}rez},
  {Luque-Escamilla}, {Paron}, \& {Miceli}}]{2016A&A...592A.125C}
{Combi}, J.~A., {Garc{\'{\i}}a}, F., {Su{\'a}rez}, A.~E., {et~al.} 2016, \aap,
  592, A125

\bibitem[{{Garc{\'{\i}}a-Senz} {et~al.}(2012){Garc{\'{\i}}a-Senz}, {Badenes},
  \& {Serichol}}]{2012ApJ...745...75G}
{Garc{\'{\i}}a-Senz}, D., {Badenes}, C., \& {Serichol}, N. 2012, \apj, 745, 75

\bibitem[{{Garc{\'{\i}}a-Senz} \& {Bravo}(2005)}]{2005A&A...430..585G}
{Garc{\'{\i}}a-Senz}, D. \& {Bravo}, E. 2005, \aap, 430, 585

\bibitem[{{Georgelin} {et~al.}(1988){Georgelin}, {Boulesteix}, {Georgelin}, {Le
  Coarer}, \& {Marcelin}}]{1988A&A...205...95G}
{Georgelin}, Y.~M., {Boulesteix}, J., {Georgelin}, Y.~P., {Le Coarer}, E., \&
  {Marcelin}, M. 1988, \aap, 205, 95

\bibitem[{{Guainazzi} \& {Tashiro}(2018)}]{2018arXiv180706903G}
{Guainazzi}, M. \& {Tashiro}, M.~S. 2018, ArXiv e-prints
  [\eprint[arXiv]{1807.06903}]

\bibitem[{{Hayato} {et~al.}(2010){Hayato}, {Yamaguchi}, {Tamagawa}, {Katsuda},
  {Hwang}, {Hughes}, {Ozawa}, {Bamba}, {Kinugasa}, {Terada}, {Furuzawa},
  {Kunieda}, \& {Makishima}}]{2010ApJ...725..894H}
{Hayato}, A., {Yamaguchi}, H., {Tamagawa}, T., {et~al.} 2010, \apj, 725, 894

\bibitem[{{Hitomi Collaboration} {et~al.}(2016){Hitomi Collaboration},
  {Aharonian}, {Akamatsu}, {Akimoto}, {Allen}, {Anabuki}, {Angelini}, {Arnaud},
  {Audard}, {Awaki}, {Axelsson}, {Bamba}, {Bautz}, {Blandford}, {Brenneman},
  {Brown}, {Bulbul}, {Cackett}, {Chernyakova}, {Chiao}, {Coppi}, {Costantini},
  {de Plaa}, {den Herder}, {Done}, {Dotani}, {Ebisawa}, {Eckart}, {Enoto},
  {Ezoe}, {Fabian}, {Ferrigno}, {Foster}, {Fujimoto}, {Fukazawa}, {Furuzawa},
  {Galeazzi}, {Gallo}, {Gandhi}, {Giustini}, {Goldwurm}, {Gu}, {Guainazzi},
  {Haba}, {Hagino}, {Hamaguchi}, {Harrus}, {Hatsukade}, {Hayashi}, {Hayashi},
  {Hayashida}, {Hiraga}, {Hornschemeier}, {Hoshino}, {Hughes}, {Iizuka},
  {Inoue}, {Inoue}, {Ishibashi}, {Ishida}, {Ishikawa}, {Ishisaki}, {Itoh},
  {Iyomoto}, {Kaastra}, {Kallman}, {Kamae}, {Kara}, {Kataoka}, {Katsuda},
  {Katsuta}, {Kawaharada}, {Kawai}, {Kelley}, {Khangulyan}, {Kilbourne},
  {King}, {Kitaguchi}, {Kitamoto}, {Kitayama}, {Kohmura}, {Kokubun}, {Koyama},
  {Koyama}, {Kretschmar}, {Krimm}, {Kubota}, {Kunieda}, {Laurent}, {Lebrun},
  {Lee}, {Leutenegger}, {Limousin}, {Loewenstein}, {Long}, {Lumb}, {Madejski},
  {Maeda}, {Maier}, {Makishima}, {Markevitch}, {Matsumoto}, {Matsushita},
  {McCammon}, {McNamara}, {Mehdipour}, {Miller}, {Miller}, {Mineshige},
  {Mitsuda}, {Mitsuishi}, {Miyazawa}, {Mizuno}, {Mori}, {Mori}, {Moseley},
  {Mukai}, {Murakami}, {Murakami}, {Mushotzky}, {Nagino}, {Nakagawa},
  {Nakajima}, {Nakamori}, {Nakano}, {Nakashima}, {Nakazawa}, {Nobukawa},
  {Noda}, {Nomachi}, {O'Dell}, {Odaka}, {Ohashi}, {Ohno}, {Okajima}, {Ota},
  {Ozaki}, {Paerels}, {Paltani}, {Parmar}, {Petre}, {Pinto}, {Pohl}, {Porter},
  {Pottschmidt}, {Ramsey}, {Reynolds}, {Russell}, {Safi-Harb}, {Saito},
  {Sakai}, {Sameshima}, {Sato}, {Sato}, {Sato}, {Sawada}, {Schartel},
  {Serlemitsos}, {Seta}, {Shidatsu}, {Simionescu}, {Smith}, {Soong}, {Stawarz},
  {Sugawara}, {Sugita}, {Szymkowiak}, {Tajima}, {Takahashi}, {Takahashi},
  {Takeda}, {Takei}, {Tamagawa}, {Tamura}, {Tamura}, {Tanaka}, {Tanaka},
  {Tanaka}, {Tashiro}, {Tawara}, {Terada}, {Terashima}, {Tombesi}, {Tomida},
  {Tsuboi}, {Tsujimoto}, {Tsunemi}, {Tsuru}, {Uchida}, {Uchiyama}, {Uchiyama},
  {Ueda}, {Ueda}, {Ueno}, {Uno}, {Urry}, {Ursino}, {de Vries}, {Watanabe},
  {Werner}, {Wik}, {Wilkins}, {Williams}, {Yamada}, {Yamaguchi}, {Yamaoka},
  {Yamasaki}, {Yamauchi}, {Yamauchi}, {Yaqoob}, {Yatsu}, {Yonetoku}, {Yoshida},
  {Yuasa}, {Zhuravleva}, \& {Zoghbi}}]{2016Natur.535..117H}
{Hitomi Collaboration}, {Aharonian}, F., {Akamatsu}, H., {et~al.} 2016, \nat,
  535, 117

\bibitem[{{Hitomi Collaboration} {et~al.}(2017){Hitomi Collaboration},
  {Aharonian}, {Akamatsu}, {Akimoto}, {Allen}, {Angelini}, {Audard}, {Awaki},
  {Axelsson}, {Bamba}, {Bautz}, {Blandford}, {Brenneman}, {Brown}, {Bulbul},
  {Cackett}, {Chernyakova}, {Chiao}, {Coppi}, {Costantini}, {de Plaa}, {den
  Herder}, {Done}, {Dotani}, {Ebisawa}, {Eckart}, {Enoto}, {Ezoe}, {Fabian},
  {Ferrigno}, {Foster}, {Fujimoto}, {Fukazawa}, {Furuzawa}, {Galeazzi},
  {Gallo}, {Gandhi}, {Giustini}, {Goldwurm}, {Gu}, {Guainazzi}, {Haba},
  {Hagino}, {Hamaguchi}, {Harrus}, {Hatsukade}, {Hayashi}, {Hayashi},
  {Hayashida}, {Hiraga}, {Hornschemeier}, {Hoshino}, {Hughes}, {Ichinohe},
  {Iizuka}, {Inoue}, {Inoue}, {Ishida}, {Ishikawa}, {Ishisaki}, {Iwai},
  {Kaastra}, {Kallman}, {Kamae}, {Kataoka}, {Katsuda}, {Kawai}, {Kelley},
  {Kilbourne}, {Kitaguchi}, {Kitamoto}, {Kitayama}, {Kohmura}, {Kokubun},
  {Koyama}, {Koyama}, {Kretschmar}, {Krimm}, {Kubota}, {Kunieda}, {Laurent},
  {Lee}, {Leutenegger}, {Limousine}, {Loewenstein}, {Long}, {Lumb}, {Madejski},
  {Maeda}, {Maier}, {Makishima}, {Markevitch}, {Matsumoto}, {Matsushita},
  {McCammon}, {McNamara}, {Mehdipour}, {Miller}, {Miller}, {Mineshige},
  {Mitsuda}, {Mitsuishi}, {Miyazawa}, {Mizuno}, {Mori}, {Mori}, {Mukai},
  {Murakami}, {Mushotzky}, {Nakagawa}, {Nakajima}, {Nakamori}, {Nakashima},
  {Nakazawa}, {Nobukawa}, {Nobukawa}, {Noda}, {Odaka}, {Ohashi}, {Ohno},
  {Okajima}, {Ota}, {Ozaki}, {Paerels}, {Paltani}, {Petre}, {Pinto}, {Porter},
  {Pottschmidt}, {Reynolds}, {Safi-Harb}, {Saito}, {Sakai}, {Sasaki}, {Sato},
  {Sato}, {Sato}, {Sawada}, {Schartel}, {Serlemitsos}, {Seta}, {Shidatsu},
  {Simionescu}, {Smith}, {Soong}, {Stawarz}, {Sugawara}, {Sugita},
  {Szymkowiak}, {Tajima}, {Takahashi}, {Takahashi}, {Takeda}, {Takei},
  {Tamagawa}, {Tamura}, {Tanaka}, {Tanaka}, {Tanaka}, {Tashiro}, {Tawara},
  {Terada}, {Terashima}, {Tombesi}, {Tomida}, {Tsuboi}, {Tsujimoto}, {Tsunemi},
  {Go Tsuru}, {Uchida}, {Uchiyama}, {Uchiyama}, {Ueda}, {Ueda}, {Uno}, {Urry},
  {Ursino}, {de Vries}, {Watanabe}, {Werner}, {Wik}, {Wilkins}, {Williams},
  {Yamada}, {Yamaguchi}, {Yamaoka}, {Yamasaki}, {Yamauchi}, {Yamauchi},
  {Yaqoob}, {Yatsu}, {Yonetoku}, {Zhuravleva}, \&
  {Zoghbi}}]{2017Natur.551..478H}
{Hitomi Collaboration}, {Aharonian}, F., {Akamatsu}, H., {et~al.} 2017, \nat,
  551, 478

\bibitem[{{Hitomi Collaboration} {et~al.}(2018{\natexlab{a}}){Hitomi
  Collaboration}, {Aharonian}, {Akamatsu}, {Akimoto}, {Allen}, {Angelini},
  {Audard}, {Awaki}, {Axelsson}, {Bamba}, {Bautz}, {Blandford}, {Brenneman},
  {Brown}, {Bulbul}, {Cackett}, {Chernyakova}, {Chiao}, {Coppi}, {Costantini},
  {de Plaa}, {de Vries}, {den Herder}, {Done}, {Dotani}, {Ebisawa}, {Eckart},
  {Enoto}, {Ezoe}, {Fabian}, {Ferrigno}, {Foster}, {Fujimoto}, {Fukazawa},
  {Furuzawa}, {Galeazzi}, {Gallo}, {Gandhi}, {Giustini}, {Goldwurm}, {Gu},
  {Guainazzi}, {Haba}, {Hagino}, {Hamaguchi}, {Harrus}, {Hatsukade}, {Hayashi},
  {Hayashi}, {Hayashida}, {Hell}, {Hiraga}, {Hornschemeier}, {Hoshino},
  {Hughes}, {Ichinohe}, {Iizuka}, {Inoue}, {Inoue}, {Ishida}, {Ishikawa},
  {Ishisaki}, {Iwai}, {Kaastra}, {Kallman}, {Kamae}, {Kataoka}, {Katsuda},
  {Kawai}, {Kelley}, {Kilbourne}, {Kitaguchi}, {Kitamoto}, {Kitayama},
  {Kohmura}, {Kokubun}, {Koyama}, {Koyama}, {Kretschmar}, {Krimm}, {Kubota},
  {Kunieda}, {Laurent}, {Lee}, {Leutenegger}, {Limousin}, {Loewenstein},
  {Long}, {Lumb}, {Madejski}, {Maeda}, {Maier}, {Makishima}, {Markevitch},
  {Matsumoto}, {Matsushita}, {McCammon}, {McNamara}, {Mehdipour}, {Miller},
  {Miller}, {Mineshige}, {Mitsuda}, {Mitsuishi}, {Miyazawa}, {Mizuno}, {Mori},
  {Mori}, {Mukai}, {Murakami}, {Mushotzky}, {Nakagawa}, {Nakajima}, {Nakamori},
  {Nakashima}, {Nakazawa}, {Nobukawa}, {Nobukawa}, {Noda}, {Odaka}, {Ohashi},
  {Ohno}, {Okajima}, {Ota}, {Ozaki}, {Paerels}, {Paltani}, {Petre}, {Pinto},
  {Porter}, {Pottschmidt}, {Reynolds}, {Safi-Harb}, {Saito}, {Sakai}, {Sasaki},
  {Sato}, {Sato}, {Sato}, {Sawada}, {Schartel}, {Serlemtsos}, {Seta},
  {Shidatsu}, {Simionescu}, {Smith}, {Soong}, {Stawarz}, {Sugawara}, {Sugita},
  {Szymkowiak}, {Tajima}, {Takahashi}, {Takahashi}, {Takeda}, {Takei},
  {Tamagawa}, {Tamura}, {Tanaka}, {Tanaka}, {Tanaka}, {Tashiro}, {Tawara},
  {Terada}, {Terashima}, {Tombesi}, {Tomida}, {Tsuboi}, {Tsujimoto}, {Tsunemi},
  {Tsuru}, {Uchida}, {Uchiyama}, {Uchiyama}, {Ueda}, {Ueda}, {Uno}, {Urry},
  {Ursino}, {Watanabe}, {Werner}, {Wilkins}, {Williams}, {Yamada}, {Yamaguchi},
  {Yamaoka}, {Yamasaki}, {Yamauchi}, {Yamauchi}, {Yaqoob}, {Yatsu}, {Yonetoku},
  {Zhuravleva}, {Zoghbi}, \& {Raassen}}]{2018PASJ...70...12H}
{Hitomi Collaboration}, {Aharonian}, F., {Akamatsu}, H., {et~al.}
  2018{\natexlab{a}}, \pasj, 70, 12

\bibitem[{{Hitomi Collaboration} {et~al.}(2018{\natexlab{b}}){Hitomi
  Collaboration}, {Aharonian}, {Akamatsu}, {Akimoto}, {Allen}, {Angelini},
  {Audard}, {Awaki}, {Axelsson}, {Bamba}, {Bautz}, {Blandford}, {Brenneman},
  {Brown}, {Bulbul}, {Cackett}, {Chernyakova}, {Chiao}, {Coppi}, {Costantini},
  {de Plaa}, {de Vries}, {den Herder}, {Done}, {Dotani}, {Ebisawa}, {Eckart},
  {Enoto}, {Ezoe}, {Fabian}, {Ferrigno}, {Foster}, {Fujimoto}, {Fukazawa},
  {Furuzawa}, {Galeazzi}, {Gallo}, {Gandhi}, {Giustini}, {Goldwurm}, {Gu},
  {Guainazzi}, {Haba}, {Hagino}, {Hamaguchi}, {Harrus}, {Hatsukade}, {Hayashi},
  {Hayashi}, {Hayashida}, {Hiraga}, {Hornschemeier}, {Hoshino}, {Hughes},
  {Ichinohe}, {Iizuka}, {Inoue}, {Inoue}, {Ishida}, {Ishikawa}, {Ishisaki},
  {Iwai}, {Kaastra}, {Kallman}, {Kamae}, {Kataoka}, {Katsuda}, {Kawai},
  {Kelley}, {Kilbourne}, {Kitaguchi}, {Kitamoto}, {Kitayama}, {Kohmura},
  {Kokubun}, {Koyama}, {Koyama}, {Kretschmar}, {Krimm}, {Kubota}, {Kunieda},
  {Laurent}, {Lee}, {Leutenegger}, {Limousin}, {Loewenstein}, {Long}, {Lumb},
  {Madejski}, {Maeda}, {Maier}, {Makishima}, {Markevitch}, {Matsumoto},
  {Matsushita}, {McCammon}, {McNamara}, {Mehdipour}, {Miller}, {Miller},
  {Mineshige}, {Mitsuda}, {Mitsuishi}, {Miyazawa}, {Mizuno}, {Mori}, {Mori},
  {Mukai}, {Murakami}, {Mushotzky}, {Nakagawa}, {Nakajima}, {Nakamori},
  {Nakashima}, {Nakazawa}, {Nobukawa}, {Nobukawa}, {Noda}, {Odaka}, {Ohashi},
  {Ohno}, {Okajima}, {Ota}, {Ozaki}, {Paerels}, {Paltani}, {Petre}, {Pinto},
  {Porter}, {Pottschmidt}, {Reynolds}, {Safi-Harb}, {Saito}, {Sakai}, {Sasaki},
  {Sato}, {Sato}, {Sato}, {Sato}, {Sawada}, {Schartel}, {Serlemtsos}, {Seta},
  {Shidatsu}, {Simionescu}, {Smith}, {Soong}, {Stawarz}, {Sugawara}, {Sugita},
  {Szymkowiak}, {Tajima}, {Takahashi}, {Takahashi}, {Takeda}, {Takei},
  {Tamagawa}, {Tamura}, {Tanaka}, {Tanaka}, {Tanaka}, {Tashiro}, {Tawara},
  {Terada}, {Terashima}, {Tombesi}, {Tomida}, {Tsuboi}, {Tsujimoto}, {Tsunemi},
  {Tsuru}, {Uchida}, {Uchiyama}, {Uchiyama}, {Ueda}, {Ueda}, {Uno}, {Urry},
  {Ursino}, {Watanabe}, {Werner}, {Wilkins}, {Williams}, {Yamada}, {Yamaguchi},
  {Yamaoka}, {Yamasaki}, {Yamauchi}, {Yamauchi}, {Yaqoob}, {Yatsu}, {Yonetoku},
  {Zhuravleva}, \& {Zoghbi}}]{2018PASJ...70...16H}
{Hitomi Collaboration}, {Aharonian}, F., {Akamatsu}, H., {et~al.}
  2018{\natexlab{b}}, \pasj, 70, 16

\bibitem[{{Hou} \& {Han}(2014)}]{2014A&A...569A.125H}
{Hou}, L.~G. \& {Han}, J.~L. 2014, \aap, 569, A125

\bibitem[{{Ishisaki} {et~al.}(2007){Ishisaki}, {Maeda}, {Fujimoto}, {Ozaki},
  {Ebisawa}, {Takahashi}, {Ueda}, {Ogasaka}, {Ptak}, {Mukai}, {Hamaguchi},
  {Hirayama}, {Kotani}, {Kubo}, {Shibata}, {Ebara}, {Furuzawa}, {Iizuka},
  {Inoue}, {Mori}, {Okada}, {Yokoyama}, {Matsumoto}, {Nakajima}, {Yamaguchi},
  {Anabuki}, {Tawa}, {Nagai}, {Katsuda}, {Hayashida}, {Bamba}, {Miller},
  {Sato}, \& {Yamasaki}}]{2007PASJ...59S.113I}
{Ishisaki}, Y., {Maeda}, Y., {Fujimoto}, R., {et~al.} 2007, \pasj, 59, 113

\bibitem[{{Iwamoto} {et~al.}(1999){Iwamoto}, {Brachwitz}, {Nomoto},
  {Kishimoto}, {Umeda}, {Hix}, \& {Thielemann}}]{1999ApJS..125..439I}
{Iwamoto}, K., {Brachwitz}, F., {Nomoto}, K., {et~al.} 1999, \apjs, 125, 439

\bibitem[{{Kaastra} \& {Mewe}(1993)}]{1993A&AS...97..443K}
{Kaastra}, J.~S. \& {Mewe}, R. 1993, \aaps, 97, 443

\bibitem[{{Kalberla} \& {Haud}(2015)}]{2015A&A...578A..78K}
{Kalberla}, P.~M.~W. \& {Haud}, U. 2015, \aap, 578, A78

\bibitem[{{Kamitsukasa} {et~al.}(2016){Kamitsukasa}, {Koyama}, {Nakajima},
  {Hayashida}, {Mori}, {Katsuda}, {Uchida}, \& {Tsunemi}}]{2016PASJ...68S...7K}
{Kamitsukasa}, F., {Koyama}, K., {Nakajima}, H., {et~al.} 2016, \pasj, 68, S7

\bibitem[{{Karpova} {et~al.}(2016){Karpova}, {Shternin}, {Zyuzin}, {Danilenko},
  \& {Shibanov}}]{2016MNRAS.462.3845K}
{Karpova}, A., {Shternin}, P., {Zyuzin}, D., {Danilenko}, A., \& {Shibanov}, Y.
  2016, \mnras, 462, 3845

\bibitem[{{Katsuda} {et~al.}(2015){Katsuda}, {Mori}, {Maeda}, {Tanaka},
  {Koyama}, {Tsunemi}, {Nakajima}, {Maeda}, {Ozaki}, \&
  {Petre}}]{2015ApJ...808...49K}
{Katsuda}, S., {Mori}, K., {Maeda}, K., {et~al.} 2015, \apj, 808, 49

\bibitem[{{Kosenko} {et~al.}(2010){Kosenko}, {Helder}, \&
  {Vink}}]{2010A&A...519A..11K}
{Kosenko}, D., {Helder}, E.~A., \& {Vink}, J. 2010, \aap, 519, A11

\bibitem[{{Koyama} {et~al.}(2007){Koyama}, {Tsunemi}, {Dotani}, {Bautz},
  {Hayashida}, {Tsuru}, {Matsumoto}, {Ogawara}, {Ricker}, {Doty}, {Kissel},
  {Foster}, {Nakajima}, {Yamaguchi}, {Mori}, {Sakano}, {Hamaguchi},
  {Nishiuchi}, {Miyata}, {Torii}, {Namiki}, {Katsuda}, {Matsuura}, {Miyauchi},
  {Anabuki}, {Tawa}, {Ozaki}, {Murakami}, {Maeda}, {Ichikawa}, {Prigozhin},
  {Boughan}, {Lamarr}, {Miller}, {Burke}, {Gregory}, {Pillsbury}, {Bamba},
  {Hiraga}, {Senda}, {Katayama}, {Kitamoto}, {Tsujimoto}, {Kohmura}, {Tsuboi},
  \& {Awaki}}]{2007PASJ...59S..23K}
{Koyama}, K., {Tsunemi}, H., {Dotani}, T., {et~al.} 2007, \pasj, 59, 23

\bibitem[{{Leutenegger} {et~al.}(2016){Leutenegger}, {Audard}, {Boyce},
  {Brown}, {Chiao}, {Eckart}, {Fujimoto}, {Furuzawa}, {Guainazzi}, {Haas}, {den
  Herder}, {Hayashi}, {Iizuka}, {Ishida}, {Ishisaki}, {Kelley}, {Kikuchi},
  {Kilbourne}, {Koyama}, {Kurashima}, {Maeda}, {Markevitch}, {McCammon},
  {Mitsuda}, {Mori}, {Nakaniwa}, {Okajima}, {Paltani}, {Petre}, {Porter},
  {Sato}, {Sato}, {Sawada}, {Serlemitsos}, {Seta}, {Sneiderman}, {Soong},
  {Sugita}, {Szymkowiak}, {Takei}, {Tashiro}, {Tawara}, {Tsujimoto}, {de
  Vries}, {Watanabe}, {Yamada}, \& {Yamasaki}}]{2016SPIE.9905E..3UL}
{Leutenegger}, M.~A., {Audard}, M., {Boyce}, K.~R., {et~al.} 2016, in
  \procspie, Vol. 9905, Space Telescopes and Instrumentation 2016: Ultraviolet
  to Gamma Ray, 99053U

\bibitem[{{Levine} {et~al.}(2006){Levine}, {Blitz}, \&
  {Heiles}}]{2006Sci...312.1773L}
{Levine}, E.~S., {Blitz}, L., \& {Heiles}, C. 2006, Science, 312, 1773

\bibitem[{{Lopez} {et~al.}(2009){Lopez}, {Ramirez-Ruiz}, {Badenes},
  {Huppenkothen}, {Jeltema}, \& {Pooley}}]{2009ApJ...706L.106L}
{Lopez}, L.~A., {Ramirez-Ruiz}, E., {Badenes}, C., {et~al.} 2009, \apjl, 706,
  L106

\bibitem[{{Lopez} {et~al.}(2011){Lopez}, {Ramirez-Ruiz}, {Huppenkothen},
  {Badenes}, \& {Pooley}}]{2011ApJ...732..114L}
{Lopez}, L.~A., {Ramirez-Ruiz}, E., {Huppenkothen}, D., {Badenes}, C., \&
  {Pooley}, D.~A. 2011, \apj, 732, 114

\bibitem[{{Maggi} \& {Acero}(2017)}]{2017A&A...597A..65M}
{Maggi}, P. \& {Acero}, F. 2017, \aap, 597, A65

\bibitem[{{Miller} {et~al.}(2011){Miller}, {Reynolds}, {Maitra}, {Gultekin},
  {Gehrels}, {Kennea}, {Siegel}, {Gelbord}, \& {Kuin}}]{2011ATel.3415....1M}
{Miller}, J.~M., {Reynolds}, M.~R., {Maitra}, D., {et~al.} 2011, The
  Astronomer's Telegram, 3415

\bibitem[{{Mitsuda} {et~al.}(2007){Mitsuda}, {Bautz}, {Inoue}, {Kelley},
  {Koyama}, {Kunieda}, {Makishima}, {Ogawara}, {Petre}, {Takahashi}, {Tsunemi},
  {White}, {Anabuki}, {Angelini}, {Arnaud}, {Awaki}, {Bamba}, {Boyce}, {Brown},
  {Chan}, {Cottam}, {Dotani}, {Doty}, {Ebisawa}, {Ezoe}, {Fabian}, {Figueroa},
  {Fujimoto}, {Fukazawa}, {Furusho}, {Furuzawa}, {Gendreau}, {Griffiths},
  {Haba}, {Hamaguchi}, {Harrus}, {Hasinger}, {Hatsukade}, {Hayashida}, {Henry},
  {Hiraga}, {Holt}, {Hornschemeier}, {Hughes}, {Hwang}, {Ishida}, {Ishisaki},
  {Isobe}, {Itoh}, {Iyomoto}, {Kahn}, {Kamae}, {Katagiri}, {Kataoka},
  {Katayama}, {Kawai}, {Kilbourne}, {Kinugasa}, {Kissel}, {Kitamoto}, {Kohama},
  {Kohmura}, {Kokubun}, {Kotani}, {Kotoku}, {Kubota}, {Madejski}, {Maeda},
  {Makino}, {Markowitz}, {Matsumoto}, {Matsumoto}, {Matsuoka}, {Matsushita},
  {McCammon}, {Mihara}, {Misaki}, {Miyata}, {Mizuno}, {Mori}, {Mori}, {Morii},
  {Moseley}, {Mukai}, {Murakami}, {Murakami}, {Mushotzky}, {Nagase}, {Namiki},
  {Negoro}, {Nakazawa}, {Nousek}, {Okajima}, {Ogasaka}, {Ohashi}, {Oshima},
  {Ota}, {Ozaki}, {Ozawa}, {Parmar}, {Pence}, {Porter}, {Reeves}, {Ricker},
  {Sakurai}, {Sanders}, {Senda}, {Serlemitsos}, {Shibata}, {Soong}, {Smith},
  {Suzuki}, {Szymkowiak}, {Takahashi}, {Tamagawa}, {Tamura}, {Tamura},
  {Tanaka}, {Tashiro}, {Tawara}, {Terada}, {Terashima}, {Tomida}, {Torii},
  {Tsuboi}, {Tsujimoto}, {Tsuru}, {Turner}, {Ueda}, {Ueno}, {Ueno}, {Uno},
  {Urata}, {Watanabe}, {Yamamoto}, {Yamaoka}, {Yamasaki}, {Yamashita},
  {Yamauchi}, {Yamauchi}, {Yaqoob}, {Yonetoku}, \&
  {Yoshida}}]{2007PASJ...59S...1M}
{Mitsuda}, K., {Bautz}, M., {Inoue}, H., {et~al.} 2007, \pasj, 59, S1

\bibitem[{{Mitsuda} {et~al.}(2014){Mitsuda}, {Kelley}, {Akamatsu}, {Bialas},
  {Boyce}, {Brown}, {Canavan}, {Chiao}, {Costantini}, {den Herder}, {de Vries},
  {DiPirro}, {Eckart}, {Ezoe}, {Fujimoto}, {Haas}, {Hoshino}, {Ishikawa},
  {Ishisaki}, {Iyomoto}, {Kilbourne}, {Kimball}, {Kitamoto}, {Konami},
  {Leutenegger}, {McCammon}, {Miko}, {Mitsuishi}, {Murakami}, {Murakami},
  {Noda}, {Ogawa}, {Ohashi}, {Okamoto}, {Ota}, {Paltani}, {Porter}, {Sato},
  {Sato}, {Sawada}, {Seta}, {Shinozaki}, {Shirron}, {Sneiderman}, {Sugita},
  {Szymkowiak}, {Takei}, {Tamagawa}, {Tashiro}, {Terada}, {Tsujimoto},
  {Yamada}, \& {Yamasaki}}]{2014SPIE.9144E..2AM}
{Mitsuda}, K., {Kelley}, R.~L., {Akamatsu}, H., {et~al.} 2014, in \procspie,
  Vol. 9144, Space Telescopes and Instrumentation 2014: Ultraviolet to Gamma
  Ray, 91442A

\bibitem[{{Nakanishi} \& {Sofue}(2016)}]{2016PASJ...68....5N}
{Nakanishi}, H. \& {Sofue}, Y. 2016, \pasj, 68, 5

\bibitem[{{Parker} {et~al.}(2005){Parker}, {Phillipps}, {Pierce}, {Hartley},
  {Hambly}, {Read}, {MacGillivray}, {Tritton}, {Cass}, {Cannon}, {Cohen},
  {Drew}, {Frew}, {Hopewell}, {Mader}, {Malin}, {Masheder}, {Morgan}, {Morris},
  {Russeil}, {Russell}, \& {Walker}}]{2005MNRAS.362..689P}
{Parker}, Q.~A., {Phillipps}, S., {Pierce}, M.~J., {et~al.} 2005, \mnras, 362,
  689

\bibitem[{{Reid}(1993)}]{1993ARA&A..31..345R}
{Reid}, M.~J. 1993, \araa, 31, 345

\bibitem[{{Reynolds} {et~al.}(2013){Reynolds}, {Loi}, {Murphy}, {Miller},
  {Maitra}, {G{\"u}ltekin}, {Gehrels}, {Kennea}, {Siegel}, {Gelbord}, {Kuin},
  {Moss}, {Reeves}, {Robbins}, {Gaensler}, {Reis}, \&
  {Petre}}]{2013ApJ...766..112R}
{Reynolds}, M.~T., {Loi}, S.~T., {Murphy}, T., {et~al.} 2013, \apj, 766, 112

\bibitem[{{Sato} {et~al.}(2016){Sato}, {Koyama}, {Lee}, \&
  {Takahashi}}]{2016PASJ...68S...8S}
{Sato}, T., {Koyama}, K., {Lee}, S.-H., \& {Takahashi}, T. 2016, \pasj, 68, S8

\bibitem[{{Sawada} \& {Koyama}(2012)}]{2012PASJ...64...81S}
{Sawada}, M. \& {Koyama}, K. 2012, \pasj, 64, 81

\bibitem[{{Sawada} {et~al.}(2012){Sawada}, {Nakashima}, {Nobukawa}, {Uchiyama},
  \& {XIS Team}}]{2012AIPC.1427..245S}
{Sawada}, M., {Nakashima}, S., {Nobukawa}, M., {Uchiyama}, H., \& {XIS Team}.
  2012, in American Institute of Physics Conference Series, Vol. 1427, American
  Institute of Physics Conference Series, ed. R.~{Petre}, K.~{Mitsuda}, \&
  L.~{Angelini}, 245--246

\bibitem[{{Serlemitsos} {et~al.}(2007){Serlemitsos}, {Soong}, {Chan},
  {Okajima}, {Lehan}, {Maeda}, {Itoh}, {Mori}, {Iizuka}, {Itoh}, {Inoue},
  {Okada}, {Yokoyama}, {Itoh}, {Ebara}, {Nakamura}, {Suzuki}, {Ishida},
  {Hayakawa}, {Inoue}, {Okuma}, {Kubota}, {Suzuki}, {Osawa}, {Yamashita},
  {Kunieda}, {Tawara}, {Ogasaka}, {Furuzawa}, {Tamura}, {Shibata}, {Haba},
  {Naitou}, \& {Misaki}}]{2007PASJ...59S...9S}
{Serlemitsos}, P.~J., {Soong}, Y., {Chan}, K.-W., {et~al.} 2007, \pasj, 59, S9

\bibitem[{{Sezer} {et~al.}(2017){Sezer}, {Ergin}, \&
  {Yamazaki}}]{2017MNRAS.466.3434S}
{Sezer}, A., {Ergin}, T., \& {Yamazaki}, R. 2017, \mnras, 466, 3434

\bibitem[{{Someya} {et~al.}(2014){Someya}, {Bamba}, \&
  {Ishida}}]{2014PASJ...66...26S}
{Someya}, K., {Bamba}, A., \& {Ishida}, M. 2014, \pasj, 66, 26

\bibitem[{{Spitzer}(1978)}]{1978ppim.book.....S}
{Spitzer}, L. 1978, {Physical processes in the interstellar medium}

\bibitem[{{Takahashi} {et~al.}(2016){Takahashi}, {Kokubun}, {Mitsuda},
  {Kelley}, {Ohashi}, {Aharonian}, {Akamatsu}, {Akimoto}, {Allen}, {Anabuki},
  \& et~al.}]{2016SPIE.9905E..0UT}
{Takahashi}, T., {Kokubun}, M., {Mitsuda}, K., {et~al.} 2016, in \procspie,
  Vol. 9905, Space Telescopes and Instrumentation 2016: Ultraviolet to Gamma
  Ray, 99050U

\bibitem[{{Takata} {et~al.}(2016){Takata}, {Nobukawa}, {Uchida}, {Tsuru},
  {Tanaka}, \& {Koyama}}]{2016PASJ...68S...3T}
{Takata}, A., {Nobukawa}, M., {Uchida}, H., {et~al.} 2016, \pasj, 68, S3

\bibitem[{{Takeuchi} {et~al.}(2016){Takeuchi}, {Yamaguchi}, \&
  {Tamagawa}}]{2016PASJ...68S...9T}
{Takeuchi}, Y., {Yamaguchi}, H., \& {Tamagawa}, T. 2016, \pasj, 68, S9

\bibitem[{{Tang} \& {Chevalier}(2017)}]{2017MNRAS.465.3793T}
{Tang}, X. \& {Chevalier}, R.~A. 2017, \mnras, 465, 3793

\bibitem[{{Tawa} {et~al.}(2008){Tawa}, {Hayashida}, {Nagai}, {Nakamoto},
  {Tsunemi}, {Yamaguchi}, {Ishisaki}, {Miller}, {Mizuno}, {Dotani}, {Ozaki}, \&
  {Katayama}}]{2008PASJ...60S..11T}
{Tawa}, N., {Hayashida}, K., {Nagai}, M., {et~al.} 2008, \pasj, 60, S11

\bibitem[{{Truelove} \& {McKee}(1999)}]{1999ApJS..120..299T}
{Truelove}, J.~K. \& {McKee}, C.~F. 1999, \apjs, 120, 299

\bibitem[{{Uchida} {et~al.}(2013){Uchida}, {Yamaguchi}, \&
  {Koyama}}]{2013ApJ...771...56U}
{Uchida}, H., {Yamaguchi}, H., \& {Koyama}, K. 2013, \apj, 771, 56

\bibitem[{{Uchiyama} {et~al.}(2009){Uchiyama}, {Ozawa}, {Matsumoto}, {Tsuru},
  {Koyama}, {Kimura}, {Uchida}, {Nakajima}, {Hayashida}, {Tsunemi}, {Mori},
  {Bamba}, {Ozaki}, {Dotani}, {Takei}, {Murakami}, {Mori}, {Ishisaki},
  {Kohmura}, {Prigozhin}, {Kissel}, {Miller}, {LaMarr}, \&
  {Bautz}}]{2009PASJ...61S...9U}
{Uchiyama}, H., {Ozawa}, M., {Matsumoto}, H., {et~al.} 2009, \pasj, 61, S9

\bibitem[{{Urdampilleta} {et~al.}(2017){Urdampilleta}, {Kaastra}, \&
  {Mehdipour}}]{2017A&A...601A..85U}
{Urdampilleta}, I., {Kaastra}, J.~S., \& {Mehdipour}, M. 2017, \aap, 601, A85

\bibitem[{{Wang} \& {Chevalier}(2001)}]{2001ApJ...549.1119W}
{Wang}, C.-Y. \& {Chevalier}, R.~A. 2001, \apj, 549, 1119

\bibitem[{{Wilms} {et~al.}(2000){Wilms}, {Allen}, \&
  {McCray}}]{2000ApJ...542..914W}
{Wilms}, J., {Allen}, A., \& {McCray}, R. 2000, \apj, 542, 914

\bibitem[{{Yamaguchi} {et~al.}(2015){Yamaguchi}, {Badenes}, {Foster}, {Bravo},
  {Williams}, {Maeda}, {Nobukawa}, {Eriksen}, {Brickhouse}, {Petre}, \&
  {Koyama}}]{2015ApJ...801L..31Y}
{Yamaguchi}, H., {Badenes}, C., {Foster}, A.~R., {et~al.} 2015, \apjl, 801, L31

\bibitem[{{Yamaguchi} {et~al.}(2014{\natexlab{a}}){Yamaguchi}, {Badenes},
  {Petre}, {Nakano}, {Castro}, {Enoto}, {Hiraga}, {Hughes}, {Maeda},
  {Nobukawa}, {Safi-Harb}, {Slane}, {Smith}, \& {Uchida}}]{2014ApJ...785L..27Y}
{Yamaguchi}, H., {Badenes}, C., {Petre}, R., {et~al.} 2014{\natexlab{a}},
  \apjl, 785, L27

\bibitem[{{Yamaguchi} {et~al.}(2014{\natexlab{b}}){Yamaguchi}, {Eriksen},
  {Badenes}, {Hughes}, {Brickhouse}, {Foster}, {Patnaude}, {Petre}, {Slane}, \&
  {Smith}}]{2014ApJ...780..136Y}
{Yamaguchi}, H., {Eriksen}, K.~A., {Badenes}, C., {et~al.} 2014{\natexlab{b}},
  \apj, 780, 136

\bibitem[{{Yamaguchi} {et~al.}(2008){Yamaguchi}, {Koyama}, {Katsuda},
  {Nakajima}, {Hughes}, {Bamba}, {Hiraga}, {Mori}, {Ozaki}, \&
  {Tsuru}}]{2008PASJ...60S.141Y}
{Yamaguchi}, H., {Koyama}, K., {Katsuda}, S., {et~al.} 2008, \pasj, 60, S141

\bibitem[{{Yamaguchi} {et~al.}(2012){Yamaguchi}, {Tanaka}, {Maeda}, {Slane},
  {Foster}, {Smith}, {Katsuda}, \& {Yoshii}}]{2012ApJ...749..137Y}
{Yamaguchi}, H., {Tanaka}, M., {Maeda}, K., {et~al.} 2012, \apj, 749, 137

\bibitem[{{Zhou} \& {Vink}(2018)}]{2018A&A...615A.150Z}
{Zhou}, P. \& {Vink}, J. 2018, \aap, 615, A150

\end{thebibliography}

\end{document}